\documentstyle[12pt]{article}
\input epsf
\begin{document}
\title{
\hfill{\small IPNO/TH 94-30}
\vspace*{0.5cm}\\
\sc
Single-particle density of states, bound states, phase-shift flip, 
\\and a resonance in  the presence of an Aharonov-Bohm potential
}
\author{Alexander Moroz
\thanks{Electronic address: {\tt am@th.ph.bham.ac.uk}}
\thanks{On leave from Institute of Physics, Na Slovance 2, CZ-180 40 Praha 8, 
Czech Republic
}%
\vspace*{0.3cm}}
\date{
\protect\normalsize
\it 
 Division de Physique Th\'{e}orique\thanks{Unit\'{e} de 
Recherche des Universit\'{e}s Paris XI et Paris VI
associ\'{e}e au CNRS}, Institut de Physique Nucl\'{e}aire,\\
\it Universit\'{e} Paris-Sud, F-91 406 Orsay Cedex, France\\
and\\
${}$\thanks{Present address}\,School of Physics and Space Research, 
University of Birmingham,
Edgbaston, Birmingham B15 2TT, U. K.
}
\maketitle
\begin{center}
{\large\sc abstract}
\end{center}
Both the nonrelativistic scattering and the spectrum in the presence of 
the Aharonov-Bohm potential are analyzed and the single-particle
density of states for different self-adjoint extensions
is calculated. The single-particle density of states is shown 
to be a symmetric and periodic function of the flux which depends
only on the distance from the nearest integer.
 The Krein-Friedel formula 
for this long-ranged potential is shown to be valid
when regularized with the $\zeta$ function. The limit when the radius $R$
of the flux tube shrinks to zero is discussed.
For $R\neq 0$ and in the case of an anomalous magnetic moment 
$g_m>2$ (note, e.\ g., that $g_m=2.002\ 32$ for the electron)
the coupling for spin-down electrons is enhanced and
bound states occur in the spectrum. Their number does depend on
a regularization and generically does not match with the number
of zero modes in a given field that occur when $g_m=2$.
Provided the coupling with the interior of the flux tube is not 
renormalized to a critical one neither bound states 
nor zero modes survive the limit $R\rightarrow 0$. 
The Aharonov-Casher theorem on the
number of zero modes is corrected for the singular 
field configuration. Whenever a bound state does survive 
the $R\rightarrow 0$ limit it is always
accompanied by a resonance. The presence of a bound state
manifests itself in the asymmetric differential scattering
cross section that can give rise to the Hall effect.
The Hall resistivity is calculated in the dilute vortex limit.
The magnetic moment coupling and  not the spin is
shown to be the primary source for the phase-shift flip
that may occur even in its absence. 
The total energy of the system consisting of particles and field
is discussed.
An application to persistent currents in the plane 
for both spinless and spin one-half
fermions is given. Persistent currents are also predicted
to exist in the field of a cosmic string.
The $2nd$ virial coefficient of anyons with a 
short range $\delta$-function interaction is calculated. 
The coefficient is shown to be remarkably 
stable when such an interaction is switched on.
Several suggestions for new experiments are given.

\vspace*{0.2cm}

{\footnotesize
\noindent PACS numbers : 03.65.Bz, 03-80.+r, 05.30.-d, 73.50.-h}

\vspace*{1.6cm}

\begin{center}
{\bf (Phys. Rev. A 53, 669-694 (1996))}
\end{center}
\thispagestyle{empty}
\baselineskip 20pt
\newpage
\setcounter{page}{1}
\section{Introduction}
%
In this paper we shall report on several physical phenomena and 
calculate several quantities in the presence of 
the Aharonov-Bohm (AB) 
potential \cite{AB}, which, in the radial gauge, is given by 
\begin{equation}
A_r=0,\hspace*{1cm} 
A_\varphi=\frac{\Phi}{2\pi r}=\frac{\alpha}{2\pi r}\,\Phi_0.
\label{abpot}
\end{equation}
Usually, $\Phi=\alpha\,\Phi_0$ is the total flux through the flux tube
and  $\Phi_0$ is the flux quantum, $\Phi_0=hc/|e|$. 
The AB potential will be considered here in a more general
sense since formally the same potential (of nonmagnetic origin)
is generated around a cosmic string.
The parameter $\Phi$ is then $1/Q_{Higgs}$, $\alpha=e/Q_{Higgs}$, 
and $\Phi_0=2\pi/e$ in the units $\hbar=c=1$ with
$e$ and $Q_{Higgs}$ being respectively the charge of a test 
particle and the charge of the Higgs particle \cite{AW}.
Experimentally, the situation of an infinitely thin flux tube is realized
when a flux tube has a  radius $R$  which is negligibly small
when compared to all other length scales in the system.
Therefore, both a flux tube with a nonzero and the zero radius will
be considered. In the case when $R>0$, we shall allow generally for 
some additional interaction inside a flux tube, since such an interaction
arises, for example, in the case of the magnetic moment coupling.
The limit $R\rightarrow 0$ will then depend on the physics
inside the flux tube. In  a rigorous mathematical sense,
different physics inside the flux tube will be described by 
different Hamiltonians given as a certain self-adjoint extension of 
a formal differential operator.

First we shall concentrate on the calculation 
of the single-particle density of states (DOS).
The reason is that the DOS is the quantity of basic interest
and provides an important link 
between different physical quantities.
Knowledge of the DOS determines [via the Laplace transform, see Eq. 
(\ref{defz}) below]
the partition function $Z(\beta)$, virial coefficients, 
and in the case of the Dirac equation a relation
between effective energy, induced fermion number,
and the axial anomaly \cite{NS,AM89}. 
It has been used \cite{CMO1} to
calculate the persistent current of free electrons induced in the plane
by the AB potential \cite{AA}.
The DOS will be calculated in two different ways: 
first, directly through
the resolvent and second, by only using the scattering properties 
of the AB potential. After the DOS and differential scattering
cross sections are calculated various applications are discussed.

The plan of the paper is as follows. 
We shall start with the case of the impenetrable flux tube
when the flux tube is exterior to the system.
In Sec. \ref{sec:dos}
the basic facts about the nonrelativistic AB scattering 
in this case  are summarized. There are neither bound states
nor zero modes and the DOS is determined solely in terms
of the continuous spectrum.
The single-particle DOS is calculated there directly
from the resolvent. One finds that for integer $\alpha$
resolvents do differ by a phase factor. However, when the
arguments  coincide they are identical and they do have the same trace and
yield a DOS identical to that in free space.
We shall confirm the
anticipation of Comtet, Georgelin, and Ouvry \cite{CGO} that 
the change of the DOS is concentrated at the zero energy
[see Eq. (\ref{denreg}) below].

Different self-adjoint extensions in the case of a penetrable 
flux tube are discussed in Sec. \ref{sec:self}.
Let the coupling constant be negative, $e=-|e|$.
Then different self-adjoint extensions arise because, in the channels
$l\!=\!-n$ and $l\!=\!-n-1$, where $n$ is the integer part of
$\alpha$, there are two linearly independent square integrable at zero
solutions of an eigenvalue equation. 
These two channels are also the only two channels where a bound state 
can occur.  Bound states  are shown to parametrize   
different self-adjoint extensions and the change in the conventional
 phase shifts. Calculation of the change $\triangle_l$
is reviewed. Knowledge of the phase shift is then used 
in Sec. \ref{sec:cros} to calculate the S matrix and 
scattering cross sections. The S matrix is shown
to depend nontrivially on $\alpha$ and not to be a periodic function
of $\alpha$. Nevertheless, the S matrix gives rise to
differential and transport scattering cross sections which are
periodic functions of $\alpha$ with period $1$.
In the presence of a bound state, one finds 
in contrast to the case of the impenetrable flux tube that 
the differential scattering cross section
ceases to be symmetric with respect to the substitution
$\varphi\rightarrow -\varphi$ [see Eq.\ (\ref{dbcross})],
where $\varphi$ is a scattering angle. 
The asymmetry is easy to understand
because (for $\alpha\geq 0$) bound states are only formed 
in the channels for which $l\leq 0$.
The results are then
used in Sec. \ref{sec:krein} where the validity of the  
Krein-Friedel formula \cite{F,AMB} for the DOS
is established for a singular potential.
It is shown that the Krein-Friedel
formula when regularized with the $\zeta$ function gives the
correct DOS. This enables us to calculate the DOS
for different self-adjoint extensions. 
The  Krein-Friedel formula gives the DOS as the sum over phase
shifts [see Eq. (\ref{krein})] and  thereby  relates 
the DOS directly to the scattering properties. 
It is therefore very useful to have its extension to
singular (especially Coulomb) potentials.
One finds that whenever a bound state is present in the spectrum
it is always accompanied by a {\em resonance}
[Eq. (\ref{reso})]. 
 Rather surprisingly, the shape of the resonance 
[given in  Eq. (\ref{resshape})] is {\em not} of the 
Breit-Wigner form. Since the latter is a direct
consequence of analyticity it poses an interesting question on
the analytic structure of scattering amplitudes for singular
potentials. The existence of the {\em resonance} is quite
unexpected. It will influence the transport properties of electrons 
in the currently almost accessible experimental regime
\cite{BKP}, and the persistent current of free electrons in 
the plane \cite{CMO1}. 
In the limit of a zero energy bound state the resonance
goes to zero, too, where it  merges with the bound state into the
continuous spectrum leaving behind the phase-shift flip.

\begin{figure}
\centerline{\epsfxsize=9cm \epsfbox{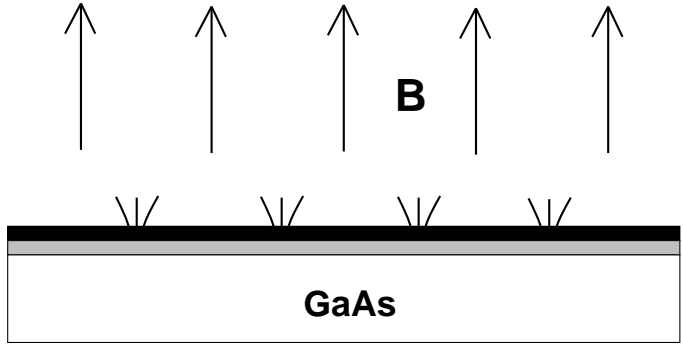}}
\caption{Black layer is a superconductor of type II put on top
of the heterostructure containing the two-dimensional electron gas
(shaded region). When this sample is put in a homogeneous magnetic
field, magnetic field  penetrates the superconductor
in Abrikosov vortices. Therefore, electrons from the heterostructure
do not move in the homogeneous magnetic field but in the field
of a penetrable flux tube.}
\label{prfgrs}
\end{figure}

In Sec. \ref{sec:reg}  the calculation of the number of bound
states for a flux tube of finite radius $R\neq 0$ is given.
The situation is considered where a short-range potential
is placed inside the flux tube. The motivation is that
vortices that can be realized in real experiments,
such as in a superconductor of type II, definitely
do have a nonzero radius. 
Another realistic realization of the penetrable
flux tube is that suggested originally by
Rammer and Shelankov \cite{RS} and later realized experimentally
by Bending, Klitzing, and Ploog \cite{BKP}, i.\ e., to put
a type II superconducting gate on top
of the heterostructure containing the two-dimensional
electron gas (2DEG) (see Fig.\ \ref{prfgrs}). 
When a magnetic field is switched on the conventional
superconductor is penetrated by vortices of flux with $\alpha=1/2$.
Therefore, electrons from the heterostructure do not move in
the homogeneous magnetic field but in the field of a 
penetrable flux tube. Electrons can penetrate to their core,
in which case the 
potential arises as the result of the magnetic moment coupling.
If the gyromagnetic ratio $g_m$ is less than $2$,
$g_m<2$,  the coupling with magnetic field is not sufficiently
strong enough to form bound states.
If the gyromagnetic ratio $g_m$ is exactly equal to $2$
(i.\ e., the magnetic moment is not anomalous)
zero modes may occur in the spectrum \cite{AC,Th}.
Their number equals, respectively, $n-1$ or $n$, where $n$ is
the integer part of $\alpha$, depending on whether the 
flux is an integer or not. There is no zero mode 
for $\alpha\leq 1$ \cite{AC,Th}.
In the region $g_m>2$, i.e., exactly where the gyromagnetic 
ratio of electron ($g_m=2.002\ 32$) lies, the coupling with 
magnetic field is enhanced and bound states do occur.
In contrast to zero modes \cite{AC} one finds that 
the number of bound states does {\em depend} on a regularization. 
For example, in the case of the cylindrical shell 
regularization \cite{Hag} their number is generally
higher than for the homogeneous field regularization 
[see  Eq. (\ref{numberb})]. The differences are attributed 
to the different energies of magnetic 
field inside the flux tube. In any case, however, 
the number of bound states does not match with the 
number of zero modes. 
The question about the existence of a bound state in
the $l\!=\!-n-1$ channel is discussed. Although generally
it is true that only one bound state is present for $0<\alpha<1$,
provided $1-\alpha$ is sufficiently small, the second bound state
does appear (cf. Ref.\ \cite{AG}).

An interpretation of different self-adjoint extensions and 
the $R\rightarrow 0$ limit are studied in Sec. \ref{sec:ren}.
The existence of a {\em critical} coupling is established.
In the case of the magnetic moment interaction with the interior
of the flux tube, the critical coupling corresponds to 
the case of the normal magnetic moment with the gyromagnetic ratio
$g_m=2$. Provided the coupling with the interior of the flux tube
is  weaker, bound states do not form at all.
If the coupling is stronger than the critical value, then, although
bound states do exist for any $R\neq 0$, they
{\em disappear} in the limit $R\rightarrow 0$.
Although it might seem surprising at first, the fact is a 
translation of the result of Berezin and Faddeev \cite{BF}
established more than $30$ years ago that the nontrivial 
$R\rightarrow 0$ limit
of a potential formally given by the Dirac delta function 
$\delta({\bf r})$ requires
a {\em renormalization} of the coupling with the interior
of the flux tube (see Ref. \cite{AGHH} for more details). 
Reference \cite{MTR} provides some other examples when one encounters
the necessity of a renormalization in the quantum 
mechanics. At first sight the renormalization
might seem to be artificial and only a mathematical obscurity.
However, one knows that the anomalous magnetic moment always
has a form factor that does depend on the energy \cite{Land}.
Therefore, in the realistic situation the renormalization
is provided by nature itself.

The point spectrum \cite{Ka} at the critical coupling is 
also considered.
One finds  that in the limit $R\rightarrow 0$ 
there are no zero modes in 
the AB potential for any $\alpha$. 
The Aharonov-Casher and the index theorems \cite{AC,Th,ASi}
are corrected in the 
sense that they do not give, respectively, the actual 
number of  zero modes in a given finite-flux 
background but rather an upper bound. In the presence of
a singular field configuration such as the AB potential,
the square integrability of solutions must also be checked
at the position of a singularity of the field. It
is shown that at such a point the square integrability fails.
Another argument showing why it happens is to note
that the only two channels
in the $R\rightarrow 0$ limit where the spectrum can differ from
the conventional one are $l\!=\!-n$ and $l\!=\!-n-1$.
However, for $R\neq 0$ zero modes [see Eq.\ (\ref{zerom})]
only occur in channels $0\geq l\geq -n+1$, and hence
they are never present in those described above.

In the presence of a bound state
the conventional phase shift (\ref{convshift}) acquires
a generally energy dependent contribution (\ref{trianl}).
In the limit $R\rightarrow 0$
bound states are possible in two different channels,
$l\!=\!-n$ and $l\!=\!-n-1$.
However, when $R\neq 0$, 
the bound state does not occur generally in the $l\!=\!-n-1$
channel. 
Therefore, it is natural to expect that the phase-shift flip
occurs generally only in the $l\!=\!-n$ channel.
According to this discussion the conditions for the occurrence 
of the phase-shift flip given by  Hagen \cite{Hag}  are necessary
but not sufficient. Moreover, since the origin of the attractive
potential inside the flux tube can be arbitrary,
our  calculation shows [Eqs. (\ref{shift}) and (\ref{intsing})]
that the phase-shift flip occurs even in the absence of the spin.
Also, in the case of particles with a spin it is not 
the spin but the magnetic moment coupling that is 
the primary source for the phase-shift flip.
A nice interpretation of the phase-shift flip appears if
the bound state energy is renormalized to zero. The phase-shift flip
then occurs as the result of merging the bound state and the
resonance into the continuous spectrum.
Although the observation of the phase shift flip is 
usually attributed
to Hagen \cite{Hag}, it was observed earlier in a complementary
situation: in the scattering off a general two-dimensional
magnetic field satisfying the finite-flux condition in the
long-wavelength limit (see Ref.\  \cite{MRW}, pp. 437 and 438).
A {\em duality} can be observed [see Eq.\ (\ref{dual})], 
important from the experimental point
of view. Imagine two flux tubes  with different
radii which are otherwise negligibly small when compared to 
other length scales in the system.
Then the scattering properties  of the flux tube
with a radius $R_1$
at momentum $k_1$ are the same as that of radius $R_2$ at momentum
$k_2$, provided that $k_1 R_1=k_2 R_2$.

Starting from the next section applications to the total energy
of the system (particles and a magnetic field),
the Hall effect, a persistent current, and the $2$nd virial
coefficients are considered. We shall show that our results are 
not only of academic but also of 
practical interest thanks to the recent developments 
in the fabrication of  microstructures and in
mesoscopic physics (see Ref.\ \cite{WW} for a recent review).
The total energy of the system  
and its stability against magnetic field creation
are discussed in Sec. \ref{sec:en}.
In general the {\em diamagnetic inequality} \cite{HSS}
tells us that that the matter is {\em stable}.
The latter was proven under the assumption of minimal
coupling. 
In our calculations we shall allow for a
speculation that the magnetic moment is
independent of momenta. We shall ignore the fact that 
the anomalous magnetic moment
has a form factor that vanishes at high momenta. 
Under these hypotheses one can show that
in the nonrelativistic case in $2+1$ dimensions,
a window may exist for the magnetic moment $g_m>2$  
in which the inequality is {\em violated}.
The reason is the formation of bound states that decouple from
the Hilbert space by taking away negative energy.
Although the dynamics, return fluxes, and the
form factor must all  play an important role in the full
(quantum-field theory) discussion of the stability,
our relation (\ref{instab}) 
is nevertheless interesting because it gives the stability
condition in terms of the ratio of the rest and  the electromagnetic 
energies.

In Sec.\ \ref{sec:hal} we shall examine consequences of
 the asymmetry of differential scattering cross sections. 
The asymmetry has important consequences as it give rise to the Hall
effect.
The Hall resistivity is then calculated in 
the dilute vortex limit [see Eq.\ (\ref{reshall})], i.\ e.,
when the multiple-scattering contribution is ignored.
In Sec. \ref{sec:pers} the results are applied to the 
persistent current of free electrons in the plane pierced 
by a flux tube. Both the spinless and the spin one-half 
cases are dicussed. The above
mentioned resemblence between the electromagnetic AB potential
and the field produced by a cosmic string \cite{AW} will enable us
to conclude that the persistent current might appear in the
latter case, too. Once again condensed matter physics gives both
the motivation and provides a test laboratory 
for a phenomenon that can occur at different
length scales in the Universe.
Finally, in Sec. \ref{sec:vir}, the $2nd$ virial 
coefficient of anyons interacting with a pairwise  interaction proportional
to the Dirac $\delta$ function is calculated.
Note that such an interaction appears in the nonrelativistic 
limit of planar field theories \cite{St,Kog}.
It will be shown that the $2$nd virial coefficients are remarkably
stable when the interaction is switched-on.
The calculation is performed directly in the continuum
without any use of the customary devices
 that makes the energy levels discrete,
such as a finite box \cite{AD} or harmonic potential
regularization \cite{CGO,BHR}. We shall show that the use of the $\zeta$-function
regularization reproduces the correct answer for the $2$nd virial 
coefficients of noninteracting anyons \cite{CGO,AD}, and generalizes 
results
of Blum et al.\ \cite{BHR} for nonrelativistic spin one-half anyons.

Note that one has the unitary equivalence between a 
spin $1/2$ charged  particle in
a two-dimensional (2D) magnetic field and a spin $1/2$ neutral 
particle with an anomalous
magnetic moment in a 2D electric field \cite{OO}.
In our presentation we confine ourselves essentially to the 
nonrelativistic Schr\"{o}dinger and Pauli equations.
The results for the Dirac and the Klein-Gordon equations,
the induced fermion number, the relation between the phase-shift
flip and the axial anomaly, and the DOS in the spacetime of a 
gravitational vortex in $2+1$ dimensions \cite{GJ}
are  discussed elsewhere \cite{AM89}. 
For the problems related to the gauge transformations 
that are not discussed here we refer to the review of 
Ruijsenaars \cite{R}.

\section{Impenetrable flux tube and the density of states}
\label{sec:dos}
Let us start with the nonrelativistic case and an 
impenetrable flux tube. We shall consider the Pauli Hamiltonian,
\begin{equation}
H= \frac{({\bf p}-\frac{e}{\hbar c}{\bf A})^2}{2m}-
\hat{\mbox{\boldmath$\mu$}}\cdot{\bf B},
\label{pauli}
\end{equation}
where $\hat{\mbox{\boldmath$\mu$}}=\mu\hat{{\bf s}}/s$ is the magnetic
moment operator, $\hat{{\bf s}}$ is the spin operator,
and $s$ is the magnitude of the particle spin.
For an electron $\mu_e =-g_m|e|\hbar/4mc=-\mu_Bg_m/2$, 
where $\mu_B$ is the Bohr
magneton and $g_m$ is the gyromagnetic ratio that characterizes 
the strength of the magnetic moment \cite{Land}. Naively $g_m=2$
but one knows that within quantum electrodynamics
 $\mu_e$ acquires radiative corrections
which depend on the fine structure constant
$\alpha_{\mbox{\protect\tiny QED}}$ ($=1/137$).
Within this framework we have 
[see Eq.\ (118.4) of Ref.\ \cite{Land}]
\begin{equation}
\mu_e=\frac{e\hbar}{2mc}\left(1+
\frac{\alpha_{\mbox{\protect\tiny  QED}}}{2\pi}-
0.328 \frac{\alpha_{\mbox{\protect\tiny  QED}}^2}{\pi^2}\right).
\end{equation}
By separating the  variables and assuming $e=-|e|$,  
the total Hamiltonian is written as a direct sum, $H=\oplus_l H_l$,
of channel radial Hamiltonians $H_l$ in the Hilbert space
$L^2[(0,\infty), rdr]$ \cite{AB,R},
\begin{equation}
H_l=-\frac{d^2}{dr^2}-\frac{1}{r}\frac{d}{dr}+\frac{\nu^2}{r^2}
+g_m\frac{\alpha}{r}\, s_z\delta(r).
\label{schrham}
\end{equation}
Here $\nu=|l+\alpha|$, $\alpha$ is
the total flux $\Phi$ in the units
of the flux quantum $\Phi_0=hc/|e|$,
and $s_z=\pm 1/2$ is the projection of the spin on the direction of the flux 
tube \cite{AB,R}.
The Schr\"{o}dinger equation is recovered upon setting $s_z=0$.

In the case of the impenetrable flux tube the spectra 
of both the Pauli and the Schr\"{o}dinger equations are identical.
Let us first consider the conventional setup where wave functions
are zero at the position of the flux tube. There are neither zero modes
nor bound states in this case as they are incompatible with the
boundary conditions. To discuss the spectrum, note that
for positive (negative) energies the eigenvalue equation in the 
$l$th channel reduces to the (modified) Bessel 
equation of the order $\nu=|l+\alpha|$,
\begin{equation}
H_l\psi_l=k^2\psi_l,
\label{eigen}
\end{equation}
with $k=\sqrt{2mE}/\hbar$. 
The boundary condition selects only regular solutions at the 
origin and the ``spectrum"  is given by
\begin{equation}
\psi_l(r,\varphi)=J_{|l+\alpha|}(kr)e^{il\varphi}.
\label{regpsi}
\end{equation}
Generically, to specify the boundary conditions at infinity
(or at zero)
is necessary only for those $l$ for which all solutions of 
(\ref{eigen}) are square integrable at infinity
(or at zero): the so-called {\em limit circle case}
(see Ref.\ \cite{RS2}, p. 152).
In general, square integrability takes the place of boundary
conditions at infinity (or at zero) if one of the 
solutions of (\ref{eigen}) {\em is not} square integrable at
infinity (or at zero): the so-called 
{\em limit point case} (see Ref.\ \cite{RS2}, p. 152). 
 From the rigorous point of view, one can speak about an 
impenetrable flux tube only provided that $\alpha$ is not an integer.
Then all wave functions $\psi_l$ (\ref{regpsi}) are zero at the origin.
Since $J_0(0)=1$, this is not the case for a flux  tube with 
an integer flux and the physics may be different from that
in free space. In what follows, we shall ignore this subtlety, 
and assume that the $l=0$ channel wave function $\psi_l$
(\ref{regpsi}) is in the spectrum, and  even in this case
we shall call the flux tube an impenetrable flux tube.

The AB potential (\ref{abpot})  is {\em long-ranged} and
the conventional phase shifts $\delta_l$'s \cite{AB},
\begin{equation}
\delta_l = \frac{1}{2}\,\pi(|l|-|l+\alpha|),
\label{convshift}
\end{equation}
are {\em singular}: they do not depend on the energy and do not
decay  to zero for $E\rightarrow\infty$. Relation (\ref{convshift})
can be intuitively understood as follows. The AB potential
creates ``vorticity'' $-\alpha$, and 
positive and negative angular momentum wave functions ``go around''
the origin, respectively, in the anticlockwise and the clockwise
directions \cite{R}.

\begin{figure}
\centerline{\epsfxsize=8cm \epsfbox{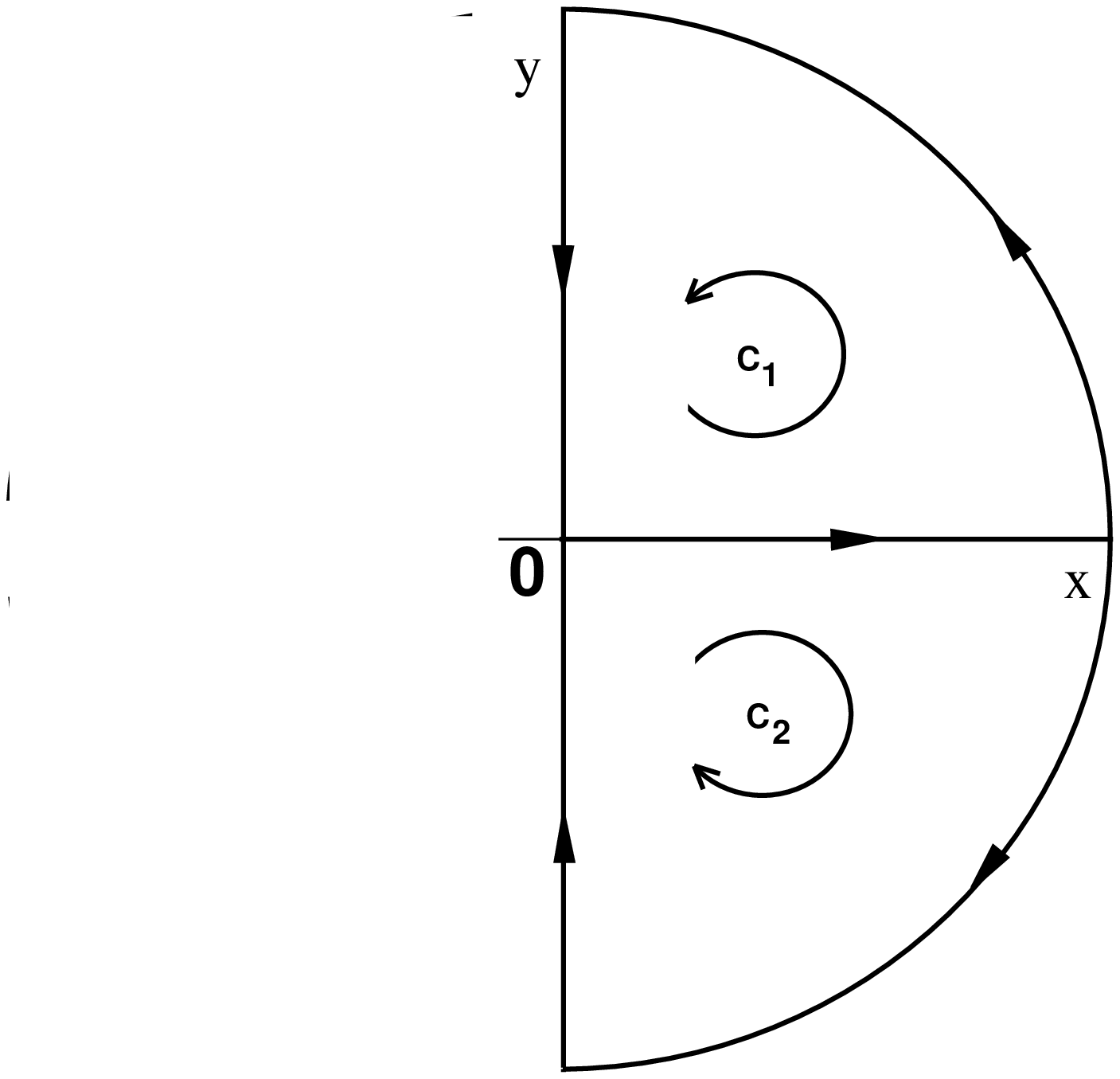}}
\caption{Contours of integration in the complex plane.}
\label{fig1}
\end{figure}

The DOS will be calculated directly from the resolvent
(the Green function)
$G_\alpha({\bf x},{\bf y},E+i\epsilon)$ according to the formula
\begin{equation}
\rho_\alpha(E)= -\frac{1}{\pi}\,\mbox{ImTr}\, G_\alpha
({\bf x},{\bf x},E+i\epsilon).
\label{resf}
\end{equation}
The integrated density of states $N_{\alpha}(E)$ is then as usual
given by
\begin{equation}
N_\alpha(E)= \int_{-\infty}^E \rho_\alpha(E')\,dE'.
\end{equation}
The eigenfunction expansion for the resolvent 
in polar coordinates ${\bf x}=(r_x,\varphi_x)$ is \cite{Pro}
\begin{equation}
G_\alpha({\bf x},{\bf y},E)=\frac{m}{\pi\hbar^2}\int_0^\infty
\frac{k dk}{q^2-k^2}\sum_{l=-\infty}^\infty
e^{il(\varphi_x-\varphi_y)}
J_{|l+\alpha|}(kr_x)J_{|l+\alpha|}(kr_y).
\label{gr:def}
\end{equation}
By using Eq.\ (\ref{resf}), one can check that the resolvent
(\ref{gr:def}) gives  the two-dimensional
free density of states $\rho_0(E)=(m/2\pi\hbar^2)V$ for 
$\alpha=n\in Z$, i.\ e., when $\alpha$ is an integer, with 
$V=\int d^2\,{\bf r}$ being the infinite volume.
In the latter case, the sum in  Eq. (\ref{gr:def}) can be 
taken exactly by means of Graf's addition theorem 
(Ref.\ \cite{AS}, relation 9.1.79),
\begin{equation}
\sum_{l=-\infty}^\infty
e^{il(\varphi_x-\varphi_y)}
J_{|l+\alpha|}(kr_x)J_{|l+\alpha|}(kr_y)
=e^{-in(\varphi_x-\varphi_y)} J_0(k|{\bf x}-{\bf y}|).
\end{equation}
By taking the integral in  Eq. (\ref{gr:def}) 
(assuming that $q^2=q^2+i\epsilon$, and $r_x<r_y$) one finds
\begin{equation}
\int_0^\infty \frac{k dk}{q^2-k^2} J_0(k|{\bf x}-{\bf y}|) =
-i\frac{\pi}{2}H^{(1)}_0(q|{\bf x}-{\bf y}|).
\label{gr:res}
\end{equation}
Therefore,
\begin{equation}
G_n({\bf x},{\bf y},E)= -i\frac{m}{2\hbar^2}\,
e^{-in(\varphi_x-\varphi_y)}H^{(1)}_0(q|{\bf x}-{\bf y}|).
\label{gr:ress}
\end{equation}
One can also perform  the integral
\begin{equation}
\int_0^\infty \frac{k dk}{q^2+i\epsilon-k^2} J_\nu(kr_x)
J_\nu(kr_y)=-\frac{\pi i}{2}J_\nu(qr_x)H^{(1)}_\nu(qr_y)
\label{int}
\end{equation}
first and then take
the remaining sum with the same result (\ref{gr:ress}).
Note that Eq.\ (\ref{int}) is also valid for nonintegral $\nu$. 
The Bessel functions are analytic functions and
the integrals in Eqs. (\ref{gr:res}) and (\ref{int}) are taken  
simply by the residue
theorem by using a suitable contour in the complex plane (see Fig.\
\ref{fig1} and Appendix \ref{app:int}).
From the `formal scattering' point of view taking the residue
at $k=|q|+i\epsilon$ corresponds to choosing outgoing
boundary conditions.
From the point of view of $L^2[(0,\infty), rdr]$ it corresponds
to taking the boundary value of the resolvent operator on 
the upper side of the cut at $[0,\infty)$ in the complex 
energy plane \cite{RR}.  Note that
\begin{equation}
{\cal G}^+_l ({\bf x},{\bf y},E)=
-i\frac{m}{2\hbar^2} \,e^{il(\varphi_x-\varphi_y)} 
J_\nu(qr_x)H^{(1)}_\nu(qr_y)
\label{presolu}
\end{equation}
is the resolvent of $H_l$ in the {\em upper-half} of 
the complex plane of $q$ \cite{RR}. 
$J_\nu(qr_x)$ is square integrable (in the measure $rdr$) 
near the origin,  and $H^{(1)}_\nu(qr_y)$ is square integrable 
near infinity. Obviously, the resolvent is not defined
for real  positive $q$ since there is no solution of the Bessel equation
that is square integrable  at infinity for such $q$ \cite{RR}
(see also asymptotic formulas in Appendix \ref{asymp}).
The point spectrum in a strict sense is empty. The spectrum
is purely continuous and lies on $[0,\infty)$ where the partial resolvent
${\cal G}^+_l$ (the Fourier transform of the retarded Green function) 
has a cut \cite{RR}.
The  partial resolvent ${\cal G}^-_l$ in the {\em lower-half} of the complex
plane (the Fourier transform of the advanced Green function) is the complex conjugate of  ${\cal G}^+_l$  \cite{RR},
\begin{equation}
{\cal G}^-_l ({\bf x},{\bf y},E)= i\frac{m}{2\hbar^2} 
\,e^{-il(\varphi_x-\varphi_y)} J_\nu(qr_x)H^{(2)}_\nu(qr_y).
\label{presoll}
\end{equation}
Note that the total Green function $G_n({\bf x},{\bf y},E)$
in the presence of an integer flux $n\neq 0$ [see Eq. (\ref{gr:ress})]
{\em differs} from
$G_0({\bf x},{\bf y},E)$ by a phase factor. However, to calculate
the DOS one needs the value of $G_n({\bf x},{\bf y},E)$
at $\varphi_x=\varphi_y$ and hence the result
for the integer flux will be the same as at $\alpha=n=0$.
The limiting value of
the resolvent operator on the lower side of the cut is
the complex conjugate of (\ref{gr:ress}), 
the discontinuity across the cut 
\begin{equation} 
G_n ({\bf x},{\bf x},E_+)-G_n
({\bf x},{\bf x},E_-)=-i(m/\hbar^2),
\end{equation}
and
\begin{equation}
-(1/\pi)\,\mbox{Im}\, G_n({\bf x},{\bf x},E_+)= m/(2\pi\hbar^2)
\end{equation}
which confirms our normalization.

Whenever $\alpha\not\in Z$  Graf's theorem  cannot be used.
To proceed further with  this case we use the fact that
(\ref{int}) has an analytic continuation on the imaginary axis
in the complex {\em momentum} plane
\begin{equation}
\int_0^\infty \frac{k dk}{q^2+k^2} J_\nu(kr_x)
J_\nu(kr_y)=\frac{\pi i}{2}J_\nu(iqr_x)H^{(1)}_\nu(iqr_y)
=I_\nu(qr_x)K_\nu(qr_y),
\end{equation}
where $I_\nu$ and $K_\nu$ are modified Bessel functions \cite{AS}.
One can obtain this result either by performing the integral directly 
or by performing the analytic continuation
in (\ref{int}).
To sum over $l$ one uses the integral representation
of these functions \cite{AS,GR}. Following the steps given in Ref.\ 
\cite{MRS} one can separate the $\alpha$-dependent contribution
\begin{eqnarray}
\lefteqn{\triangle G_\alpha({\bf x},{\bf x},E)=
G_\alpha({\bf x},{\bf x},M)-G_0({\bf x},{\bf x},M)
=
}
\nonumber\\
&&
-\frac{m}{\hbar^2}
\frac{\sin(\eta\pi)}{(2\pi)^2}
\int_{-\infty}^\infty\! d\vartheta\,\int_{-\infty}^\infty\!
d\omega e^{-M
r_x(\cosh\vartheta +\cosh\omega)}\,
\frac{e^{\eta(\vartheta-\omega)}}{1+e^{\vartheta-\omega}},
\label{grint}
\end{eqnarray}
where $M=-i\sqrt{2mE}/\hbar$ and $\eta$ is the 
nonintegral part of  $\alpha$, $0\leq \eta<1$.
After taking the trace over spatial coordinates,  
using formulas 3.512.1 and 8.334.3 of Ref.\ \cite{GR} 
(see Appendix \ref{app:int}), and returning back
to the real momentum axis,  one finally finds
\begin{equation}
\mbox{Tr}\,\triangle G_\alpha({\bf x},{\bf x},E)=
-\eta(1-\eta)\frac{1}{2E}\cdot
\end{equation}
Hence, the change 
$\triangle\rho_\alpha(E)= \rho_\alpha(E) -\rho_0(E)$
of the DOS induced by the AB potential in the {\em whole space} 
 is concentrated at zero energy
where it is proportional to the $\delta$ function,
\begin{equation}
\triangle\rho_\alpha(E)=-\frac{1}{\pi}\,\mbox{ImTr}\, 
\triangle G_\alpha ({\bf x},{\bf x},E_+)=
- \frac{1}{2}\,\eta(1-\eta)\,\delta(E).
\label{denreg}
\end{equation}
This result is similar to  the result (\ref{densb}) 
that was obtained in the 
context of anyonic physics \cite{CGO} (see also Sec. \ref{sec:vir}
below).
The same term  has been obtained
by Berry \cite{MBe} as the leading term in the semiclassical 
approximation 
to the change of the DOS for the AB (circular, heart shaped, 
and Africa shaped) quantum billiards. 
Thereby, the semiclassical approximation turns out to be exact 
in the $R\rightarrow \infty$ limit, where $R$ is 
the characteristic length of the billiards.

Let us now take the case where   $\alpha$ is negative and of the form
$\alpha=-n-\eta\leq 0$ where $n\geq 0$ is a positive integer and
$\eta$ (the fractional part of $\alpha$) satisfies $0\leq \eta<1$.
One  can  readily repeat the same procedure to obtain the change 
in the DOS. The result is exactly the same as in (\ref{denreg}),
\begin{equation}
\triangle\rho_{-\alpha}(E)= \triangle\rho_\alpha(E)=
- \frac{1}{2}\,\eta(1-\eta)\,\delta(E).
\label{denregs}
\end{equation}
Therefore, $\triangle\rho_\alpha(E)$ as a function of $\alpha$ is
a symmetric function under the transformation 
$\alpha\rightarrow -\alpha$.
Moreover, one can show that $\triangle\rho_\alpha(E)$ as a function
of $\alpha$ is only a function of a distance from 
the nearest integer.
Indeed, under the substitution
\begin{equation}
\eta'=1-\eta
\label{etasub}
\end{equation}
one finds that $\triangle\rho_\alpha(E)$ is form invariant, namely
\begin{equation}
\triangle\rho_{\alpha}(E)=- \frac{1}{2}\,\eta(1-\eta)\,\delta(E)
=- \frac{1}{2}\,\eta'(1-\eta')\,\delta(E).
\label{denform}
\end{equation}

\section{Penetrable flux tube and self-adjoint extensions}
\label{sec:self}
Physically, self-adjoint extensions in the presence of an
Aharonov-Bohm potential can be understood as follows.
In channels with  $|l+\alpha|>1$, one has a unique choice
for the resolvent of $H_l$ in $L^2[(0,\infty), rdr]$.
Apart from $J_{|l+\alpha|}(kr)$, there does not exist any other
linearly independent solution of Eq.\ (\ref{eigen})
which is square integrable at the origin. Similarly,
apart from $H_{|l+\alpha|}^{(1,2)}(kr)$, there does not exist
any other linearly independent solution of Eq.\ (\ref{eigen}) 
which is square integrable at infinity
in the upper (lower) complex half-plane of $k$. In this case, the
Hamiltonian $H_l$ is said to be in the limit point
case at both infinity and the origin \cite{RS2}, and
the square integrability takes the place of boundary conditions.
An ambiguity in defining the resolvent ${\cal G}_l^\pm$
[see Eqs.\ (\ref{presolu})
and (\ref{presoll})] of $H_l$ arises only in the channels
with $|l+\alpha|<1$, i.\ e., with $l\!=\!-n,-n-1$.
There are exactly those values of $l$ for which  Eq.\ (\ref{eigen})
has two linearly independent  (and hence all) square integrable
at the origin  solutions. These solutions can be taken to 
be  $J_{|l+\alpha|}$ and $J_{-|l+\alpha|}$.
Then, if $J_{|l+\alpha|}(qr_x)$ is replaced in either (\ref{presolu})
or (\ref{presoll}) by any linear combination of
$J_{|l+\alpha|}(qr_x)$ and $J_{-|l+\alpha|}(qr_x)$, one obtains
a well defined resolvent of $H_l$. To any particular choice of 
the linear combination corresponds a particular self-adjoint extension
of $H_l$. For these values of $l$, the AB potential 
is said to be in the limit circle case at zero
 (see Ref.\ \cite{RS2}, p. 152), and boundary conditions
at the origin must be specified to define a resolvent uniquely. 

From the formal mathematical point of view the above discussion
is repeated as follows.
Formal differential operator $H_l$  in $L^2[(0,\infty), rdr]$
is real and hence commutes with the involution operator
(complex conjugation) in the Hilbert space. Therefore,
by the von-Neumann theorem (see Ref.\ \cite{RS2}, p. 143),
 when $H_l$ is defined as a symmetric operator
on a dense set in $L^2[(0,\infty), rdr]$ 
such as $C^\infty_0(0,\infty)$
(the set of infinitely differentiable functions that 
are zero at the origin and decay exponentially at infinity)
it always can be extented to a self-adjoint operator. 
According to theorem X.11 of Ref.\ \cite{RS2}, except for the
channels with $|l+\alpha|<1$, the operators $H_l$
are already essentially self-adjoint on $C^\infty_0(0,\infty)$
[their deficiency indices are $(0,0)$].
Only the operators $H_l$   with $|l+\alpha|<1$,
i.\ e., with $l\!=\!-n,-n-1$, admit a nontrivial one-parametric family 
of self-adjoint extensions \cite{R,RR}
(their deficiency indices are $(1,1)$ \cite{R}).
Therefore, in these two cases the Hamiltonian as a self-adjoint
operator in the Hilbert space is not defined
uniquely. Different self-adjoint extensions give different 
Hamiltonians which correspond to different
physics inside the flux tube 
and different boundary conditions at its boundary
(see an example in Ref.\ \cite{RS2}, pp. 144 and 145).

To identify the particular self-adjoint extension one starts 
with the finite flux tube of radius $R$
and takes the limit $R\rightarrow 0$ (see Sec. \ref{sec:ren}).
We shall consider the situation with bound states 
of energy $E_l=-(\hbar^2/2m)\kappa_l^2$ 
in the $l\!=\!-n$ and $l\!=\!-n-1$ channels,
and with $n=[\alpha]$ denoting the nearest  integer {\em smaller} than 
or equal to $\alpha$. 
This happens when a sufficiently strong attractive short-range
potential is placed inside the flux tube. In the nonrelativistic 
case for spin-down fermions this is the case when 
their magnetic moment exceeds the value $2$ such as in  
the case of the electron which has $g_m=2.002\ 32$ \cite{AM89,BV}. 
For negative energy, the eigenvalue equation
with $H_l$ leads to the modified Bessel equation and 
the wave function is given by
\begin{equation}
B_l(r,\varphi)=K_{|l+\alpha|}(\kappa_l r)\,e^{il\varphi}.
\label{bounst}
\end{equation}
$K_{|l+\alpha|}(\kappa_l r)$
decreases exponentially for $r\rightarrow\infty$ and 
for $|l+\alpha|<1$ it is in $L^2[(0,\infty), rdr]$, although it is 
singular at the origin (see Appendix \ref{asymp}).
Nevertheless, if we want the bound state to be in the Hilbert space
the scattering states (\ref{regpsi})
in the  channels $l\!=\!-n$ and $l\!=\!-n-1$  have to be modified.
They become a combination of the 
regular and singular Bessel functions,
\begin{equation}
\psi_l(r,\varphi)=
\left[J_{|l+\alpha|}(kr)- A_lJ_{-|l+\alpha|}(kr)\right]\,e^{il\varphi}.
\label{psising}
\end{equation}
This is because $H_l$ has necessarily to be a symmetric operator. 
This means that the radial parts $\chi(r)$ of any two states 
$\psi_1({\bf r})$ and $\psi_2({\bf r})$ in the Hilbert
space have to satisfy 
\begin{equation}
r\,\left[\partial_r \chi^*_1(r)\chi_2(r) - 
\chi^*_1(r)\partial_r \chi_2(r)\right]=r\, W[\chi_1^*, \chi_2]\rightarrow 0
\label{wron}
\end{equation}
in the limit $r\rightarrow 0$,
where $W[,]$ denotes the {\em Wronskian}.
The condition (\ref{wron}) is nothing but the boundary condition 
at the origin: in the limit $r\rightarrow 0$
the logarithmic derivative $r\,\chi'(r)/\chi(r)$ of any state in the 
given Hilbert space takes a fixed value \cite{Wro}.
This is a translation of the mathematical analysis
in terms of deficiency indices \cite{R} into more physical terms.
Obviously, the Wronskian will vanish when the asymptotic
forms for $r\rightarrow 0$, up to order ${\cal O}(r)$,
of any two states are identical.
For the general scattering solution
(\ref{psising}) for $0<\nu<1$  one has (see Appendix \ref{asymp})
\begin{equation}
J_\nu(kr) -AJ_{-\nu}(kr) \sim
-\frac{A}{\Gamma(1-\nu)}\left[ \left(\frac{kr}{2}\right)^{-\nu}-
\frac{1}{A}\frac{\Gamma(1-\nu)}{\Gamma(1+\nu)}
\left(\frac{kr}{2}\right)^\nu\right]+
{\cal O}(r^{2-\nu}).
\end{equation}
The asymptotic form of (\ref{bounst}) is then determined
by that of $K_\nu(z)$
[see  Eq. (\ref{kasymp}) of Appendix \ref{asymp}]
\begin{equation}
K_\nu(z) \sim  
\frac{1}{2}\Gamma(\nu)\left[\left(\frac{z}{2}\right)^{-\nu}-
\frac{\Gamma(1-\nu)}{\Gamma(1+\nu)}
\left(\frac{z}{2}\right)^\nu\right] +{\cal O}(z^{2-\nu}).
\nonumber
\end{equation}
Therefore, the condition (\ref{wron})  
determines $A_l$ in the channels $l\!=\!-n$ and $l\!=\!-n-1$ to be 
\begin{eqnarray}
A_{-n} &=&
\left(k/\kappa_{-n}\right)^{2\eta}=
\left(E/|E_{-n}|\right)^{\eta},\nonumber\\
A_{-n-1} &=&
\left(k/\kappa_{-n-1}\right)^{2(1-\eta)}=
\left(E/|E_{-n-1}|\right)^{1-\eta}, 
\label{al}
\end{eqnarray}
i.\ e., {\em energy dependent}. 
According to (\ref{psising}) and (\ref{al}) the bound state energy
determines the spectrum of $H_l$ in the Hilbert space that
is different for different bound state energies.
Therefore, in physical terms, it is the bound state energy that
parametrizes different self-adjoint extensions. 

Quite surprisingly, the influence of the bound state on 
the wave function (\ref{psising})
is most pronounced in the limit 
$E_b\rightarrow 0$. Then  $A_l\rightarrow\infty$ 
and the regular wave function changes to the {\em singular} one,
\begin{equation}
\psi_l(r,\varphi)= J_{|l+\alpha|}(kr)\,e^{il\varphi}\rightarrow 
\psi_l(r,\varphi)= J_{-|l+\alpha|}(kr)\,e^{il\varphi}.
\label{psichan}
\end{equation}
On the other hand, in the limit of a bound state with infinite
bound energy, $A_l\rightarrow 0$ and
\begin{eqnarray}
\lefteqn{\psi_l(r,\varphi)= \left[J_{|l+\alpha|}(kr)
-  A_lJ_{-|l+\alpha|}(kr)\right]
e^{il\varphi}\rightarrow}\hspace*{2cm}
\nonumber\\
&&
\psi_l(r,\varphi)= J_{|l+\alpha|}(kr)\,e^{il\varphi}.
\label{psichan1}
\end{eqnarray}
By using the formula (9.2.5)  of Ref.\ \cite{AS}
one finds that the radial part of the general solution (\ref{psising}) 
behaves for $r\rightarrow\infty$ as follows:
\begin{equation}
R_l(r)\sim \mbox{const}\times\left(e^{-ikr}+ \frac{1-A_le^{i\pi|l+\alpha|}}
{1-A_le^{-i\pi|l+\alpha|}}e^{-i\pi(|l+\alpha|+1/2)}
e^{ikr}\right).
\end{equation}
By comparing with its behavior for $\alpha=0$ one finds
 the $l$th channel S${}_l$ matrix to be 
S${}_l=e^{2i\delta_l}$ with
\begin{equation}
\delta_l=\delta_l(E)=\frac{1}{2}\pi(|l|-|l+\alpha|)+\triangle_l(E).
\label{shift}
\end{equation}
The change $\triangle_l=\triangle_l(E)$ of the conventional phase shift
(\ref{convshift}) is determined by the equation
\begin{equation}
e^{2i\triangle_l}=\frac{1-A_le^{i\pi|l+\alpha|}}
{1-A_le^{-i\pi|l+\alpha|}},
\end{equation}
with the solution given by
\begin{equation}
\triangle_l=
\arctan\left(\frac{\sin(|l+
\alpha|\pi)}{\cos(|l+\alpha|\pi) -A_l^{-1}}\right).
\label{trianl}
\end{equation}
Here we have tacitly assumed that $A_l\rightarrow 0$ 
as $\eta\rightarrow 0$. For the discussion of this point 
see the end of Sec. \ref{sec:reg}. Note that when the bound state
energy is changed, $E_b\rightarrow E_b'$, and consequently 
$\kappa \rightarrow \kappa'$, then the phase shift $\delta_l(k)$
in the corresponding channel
at momentum $k$ equals to $\delta_l'(k')$, the phase shift
in the same channel when the bound state energy is $E_b'$,
provided that
\begin{equation}
k'=\frac{\kappa'}{\kappa}k.
\label{scaling}
\end{equation}
The energy of the bound state breaks the classical
scale invariance and sets a scale to the problem. Equation (\ref{scaling})
then expresses the scaling transformation under which the problem
is invariant. 

One sees again that the most pronounced
influence on the phase shift is in the limit $E_b\uparrow 0$.
Then $A_l^{-1}=0$,
\begin{equation}
\delta_l=\frac{1}{2}\pi(|l|+ |l+\alpha|),
\label{shflip}
\end{equation}
and one has the phase-shift flip when compared to the
conventional phase shift (\ref{convshift}).
Using the fact that phase shifts are only defined modulo $\pi$ one finds
\begin{eqnarray}
\delta_{-n}&=& \frac{1}{2}\pi(n-\eta)\simeq 
-\frac{1}{2}\pi\alpha\rightarrow 
\frac{1}{2}\pi\alpha,
\nonumber\\
\delta_{-n-1}&=& \frac{1}{2}\pi(n+\eta)
\simeq\frac{1}{2}\pi\alpha\rightarrow 
-\frac{1}{2}\pi\alpha.
\end{eqnarray}
\section{The S matrix and scattering cross sections}
\label{sec:cros}
Once phase shifts are calculated, the form of scattering amplitudes,
the S matrix, and cross sections can be discussed.
The scattering amplitude $f(k,\varphi)$ in two space dimension 
is defined as the coefficient in front of $e^{ikr}/r^{1/2}$
in the asymptotic expansions as $r\rightarrow 0$ of the
wave function (see Ref.\ \cite{RN} for a more rigorous definition),
\begin{equation}
\psi_l(r,\varphi,\varphi')\sim e^{ikr\cos(\varphi-\varphi')}+ f(k,\varphi-\varphi')\,
\frac{e^{ikr}}{r^{1/2}},
\label{scatst}
\end{equation}
where $\varphi'$ is the direction of the incident plane wave.
The scattering amplitude $f(k,\varphi)$ is related to the 
$\mbox{S}$ matrix,
\begin{equation}
(S-1)(k,\varphi)=(ik/2\pi)^{1/2} f(k,\varphi),
\label{sfrel}
\end{equation}
where $1$ stands here for the unit operator.
In the partial-wave expansion one has
\begin{equation}
f(k,\varphi)=(2\pi ik)^{-1/2}\sum_{l=-\infty}^\infty
\left(e^{2i\delta_l(k)}-1\right)
e^{il\varphi},
\end{equation}
which expresses the scattering amplitude in terms of 
the phase shifts $\delta_l(k)$.
The  differential scattering {\em cross section} is given by 
\begin{equation}
\left(\frac{d\sigma}{d\varphi}\right)(k,\varphi) =|f(k,\varphi)|^2 .
\label{difcros}
\end{equation}              

In conventional AB scattering, the phase shifts
$\delta_l$'s are given by  Eq. (\ref{convshift}).
One then finds that 
\begin{equation}
e^{2i\delta_l}=\left\{
\begin{array}{rl}
e^{-i\pi\alpha}, &l\geq -n\\
e^{i\pi\alpha},&l\leq-n-1,
\end{array}
\right.
\label{ephp}
\end{equation}
where, as above, $n$ stands for the integer part, $[\alpha]$,
of $\alpha$. Therefore, the S matrix, $s^0_\alpha(\varphi)$,
 in the AB potential is given by
\begin{eqnarray}
\lefteqn{s^0_\alpha(\varphi) 
:=\frac{1}{2\pi} \sum_{l=-\infty}^\infty e^{2i\delta_l+il\varphi}=
}
\nonumber\\
&&
\frac{\cos(\pi\alpha)}{2\pi}\sum_{l=-\infty}^\infty e^{il\varphi} +
i\frac{\sin(\pi\alpha)}{2\pi}e^{-in\varphi}
\left[\sum_{l=-\infty}^{-1} - \sum_{l=0}^\infty \right]
e^{il\varphi},
\label{sum}
\end{eqnarray}
where the superscript $0$ means
that we are dealing with conventional AB scattering.
Now, the sum in the first term in Eq. (\ref{sum}) gives
the delta function $\delta(\varphi)$.
Provided $\varphi\neq 0$ the second sum in Eq. (\ref{sum}) 
(as well as the first sum) can be taken exactly \cite{Sum} and
one finds that the term in parentheses gives
\begin{equation}
e^{-i\varphi}\frac{1}{1-e^{-i\varphi}} - \frac{1}{1-e^{i\varphi}}=
2\,\frac{1}{e^{i\varphi}-1}\cdot
\label{parterm}
\end{equation}
It is obviously singular at $\varphi=0$. Therefore, care has to
be taken to define the sum at this point. One has to resist
the temptation to extrapolate the result from those
at $\varphi\neq 0$ because the result is 
not continuous at $\varphi=0$.
One can show that the result is given by \cite{R}
\begin{equation}
s^0_\alpha(\varphi) =
\delta(\varphi)\cos\pi\alpha +i\,\frac{\sin\pi\alpha}{\pi}\,
e^{-in \varphi}\, P\frac{1}{e^{i\varphi}-1},
\label{sconv}
\end{equation}
where P denotes an analog of the principal value.
Formally,
\begin{equation}
P\,\frac{1}{e^{i\varphi}-1} =\lim_{\epsilon\downarrow 0}
\frac{1}{2}\left[
\frac{1}{e^{i\varphi}-(1-\epsilon)}
+\frac{1}{e^{i\varphi}-(1+\epsilon)}\right].
\end{equation}
The result (\ref{sconv}) can be verified by
taking the inverse Fourier transform and using the formula
\begin{equation}
P\int_{-\pi}^\pi\! d\varphi\,\frac{e^{ik\varphi}}{e^{i\varphi}-1} =
\left\{
\begin{array}{rl}
\pi, &k\geq 1\\
-\pi,&k\leq 0.
\end{array}
\right.
\end{equation}
 Note that there is nothing
particular about the $\delta$ function term in the S matrix
in the presence of an AB potential. In fact, 
$S(\varphi)=\delta(\varphi)$ is the S matrix in free space
(in the absence of any scatterer),
as can be checked directly by substituting $\alpha=0$ in Eq.\
(\ref{sum}). The $\delta$ function in Eq.\ (\ref{sconv})
simply means that a fraction  $\cos\pi\alpha$  of an incident beam
passes through the flux tube without being scattered.

In the case of negative $\alpha=-|\alpha|$ the analog of (\ref{ephp}) is
\begin{equation}
e^{2i\delta_l}=\left\{
\begin{array}{rl}
e^{i\pi|\alpha|}, &l\geq n+1\\
e^{-i\pi|\alpha|},&l\leq n.
\end{array}
\right.
\label{ephm}
\end{equation}
By repeating the above procedure leading to (\ref{sconv}), one finds
that the S matrix in the present case is
\begin{equation}
s^0_{-|\alpha|}(\varphi) =
\delta(\varphi)\cos\pi|\alpha| +i\,\frac{\sin\pi|\alpha|}{\pi}\,
e^{in \varphi}\, P\frac{1}{e^{-i\varphi}-1}\cdot
\label{sconv-}
\end{equation}
This result for $s^0_{-|\alpha|}(\varphi)$ can be verified
independently by taking the inverse Fourier transform and using
the formula
\begin{equation}
P\int_{-\pi}^\pi\! d\varphi\,\frac{e^{ik\varphi}}{e^{-i\varphi}-1} =
\left\{
\begin{array}{rl}
\pi, &k\leq -1\\
-\pi,&k\geq 0.
\end{array}
\right.
\end{equation}
A comparison of relations (\ref{sconv}) and (\ref{sconv-}) shows that,
under the transformation $\alpha\rightarrow -\alpha$,
\begin{equation}
s^0_{-|\alpha|}(\varphi)= s^0_{|\alpha|}(-\varphi).
\label{sminv}
\end{equation}

According to  Eq. (\ref{sfrel}), the scattering amplitude is given by
\begin{equation}
f_\alpha(k,\varphi)=(2\pi/ik)^{1/2} [s_\alpha-1](k,\varphi).
\label{sfrel1}
\end{equation}
For $\varphi\neq 0$ one finds
\begin{equation}
f^0_\alpha(k,\varphi)=(1/2\pi ik)^{1/2}\frac{\sin(\pi\alpha)}
{\sin(\varphi/2)}e^{-i(n+1/2)\varphi}.
\end{equation}
Hence, provided $\varphi\neq 0$, one has by using
Eq. (\ref{difcros})
\begin{equation}
\left(\frac{d\sigma^0}{d\varphi}\right)(k,\varphi) =\frac{1}{2\pi k}
\frac{\sin^2 (\pi\alpha)}{\sin^2(\varphi/2)}\cdot
\label{dcross}
\end{equation}
The total scattering cross section is infinite for $\alpha\not\in Z$ 
and vanishes for $\alpha\in Z$ \cite{R}. For the purpose 
of the experiment, note that the differential scattering 
cross section (\ref{dcross})
is {\em symmetric} under $\varphi\rightarrow-\varphi$.

When bound states are present in the $l\!=\!-n$ and in the 
$l\!=\!-n-1$ channels,
the S matrix and the scattering amplitude $f_\alpha(k,\varphi)$ are modified. 
First one finds that
\begin{equation}
s^0_\alpha\rightarrow s_\alpha=s^0_\alpha
+\triangle s^0_\alpha,
\label{sconvm}
\end{equation}
where
\begin{eqnarray}
\lefteqn{\triangle s^0_\alpha(\varphi) = \sum_{l=-n-1}^{-n}
\left[
e^{2i(\delta_{l}+\triangle_{l})}
-e^{2i\delta_{l}}\right]e^{il\varphi}
}\nonumber\\
&&
=2i\sum_{l=-n-1}^{-n}e^{2i\delta_l+i\triangle_l+il\varphi}
\sin\triangle_l.
\label{strian}
\end{eqnarray}
Here, $\triangle_l$ is the change (\ref{trianl}) of the 
conventional AB phase shift (\ref{convshift}) in the channel $l$.
Similarly, one introduces $\triangle f^0_\alpha(k,\varphi)$ as
\begin{equation}
f_\alpha(k,\varphi)= f^0_\alpha(k,\varphi)+ \triangle f^0_\alpha(k,\varphi).
\end{equation}
According to Eq. (\ref{sfrel1}) one has
\begin{equation}
\triangle f^0_\alpha(k,\varphi)=(2i/\pi k)^{1/2}
\sum_{l=-n-1}^{-n}e^{2i\delta_l+i\triangle_l+il\varphi}
\sin\triangle_l.
\label{df}
\end{equation}
Nevertheless, the property (\ref{sminv}) of the S matrix still holds
and even in the presence of bound states one has
\begin{equation}
s_{-|\alpha|}(\varphi)= s_{|\alpha|}(-\varphi).
\label{smtinv}
\end{equation}
To prove this it is sufficient to show that that a phase shift
$\delta_l(E)$ in the $l\!=\!-n$ ($l\!=\!-n-1$) channel for $\alpha\geq 0$
is the same as that in the  $l\!=\!n$ ($l\!=\!n+1$) channel 
for $\alpha\leq 0$. In the case of conventional phase shifts one can check
for this property directly from Eqs. (\ref{ephp}) and (\ref{ephm}).
Since bound state energies and $|l+\alpha|$ remain invariant under 
the transformation $\alpha\rightarrow -\alpha$ and  $l\rightarrow -l$,
according to Eq.\ (\ref{trianl}),  $\triangle_l$, and hence
the phase shift (\ref{shift}), remains invariant, too.

Note that  if $\alpha$ changes by $\pm 1$, the S matrix 
[see Eqs.\ (\ref{sconv}) and (\ref{sconvm})] as a function of $\alpha$
is not a periodic function of $\alpha$ and changes nontrivially.
This is analogous to the fact that, as was shown in Sec.\ \ref{sec:dos},
the Green's function  also
depends nontrivially on $\alpha$. However, such as in the latter case
when the Green's function gave the single-particle DOS which was periodic
in $\alpha$ with respect to the substitution
$\alpha\rightarrow \alpha \pm 1$, the S matrix will be shown to  
give scattering cross
sections that are periodic in $\alpha$ with the same period, too. 
In the absence of bound states this follows immediately from Eq.\ (\ref{dcross}). In the presence of bound states,
the differential scattering cross section for $\varphi\neq 0$
is calculated by using  Eqs. (\ref{difcros}) and (\ref{df}),
\begin{eqnarray}
\lefteqn{
\left(\frac{d\sigma}{d\varphi}\right)(k,\varphi) =
\left(\frac{d\sigma^0}{d\varphi}\right)(k,\varphi) +
\frac{8\pi}{k}\sum_{l=-n-1}^{-n}\sin^2\triangle_{l}\, +
}\nonumber\\
&&
 \frac{4}{k}\frac{\sin(\pi\alpha)}{\sin(\varphi/2)}
\left[
\sin\triangle_{-n}\cos\left(\triangle_{-n}-\pi\alpha
+\varphi/2\right) +\sin\triangle_{-n-1}\cos\left(\triangle_{-n-1}+\pi\alpha
-\varphi/2\right) \right].
\label{dbcross}
\end{eqnarray}
The periodicity of the differential cross section with respect to the substitution $\alpha\rightarrow \alpha \pm 1$ then follows from Eq.\
(\ref{dbcross}).
Note that in the presence of bound states, the differential
cross section becomes {\em asymmetric} with regard to 
$\varphi\rightarrow-\varphi$ (what is equivalent, with regard to
$\alpha\rightarrow -\alpha$). The origin of the asymmetry is 
easy to understand
as bound states for $\alpha\geq 0$ are formed only in channels 
with $l\leq 0$. 

For completeness we also give the result for the so-called
{\em transport} scattering cross section $\sigma_{tr}$,
defined by 
\begin{equation}
\sigma_{tr}:= \int_{-\pi}^\pi (1-\cos\varphi)\,
\frac{d\sigma}{d\varphi}\,d\varphi
\label{sgtr}
\end{equation}
[see Ref.\  \cite{La}, Eq.\ (139.7)]. One finds
\begin{eqnarray}
\lefteqn{\sigma_{tr} = \frac{2}{k}\,\sin^2(\pi\alpha)+
\frac{16\pi^2}{k} \sum_{l=-n-1}^{-n}
\sin^2 \triangle_l}\nonumber\\
&& -\frac{8\pi}{k} \sin(\pi\alpha)
\left[ \sin\triangle_{-n} \sin(\triangle_{-n}-\pi\alpha)-
\sin\triangle_{-n-1} \sin(\triangle_{-n-1}+\pi\alpha)
\right].
\label{sgtrres}
\end{eqnarray}
As is the differential scattering cross section,
the transport scattering cross section is also periodic
with respect to the substitution $\alpha \rightarrow \alpha\pm 1$.

\section{The Krein-Friedel formula and the resonance}
\label{sec:krein}
Now, we shall show that to calculate the change of the  
{\em integrated} density of states (IDOS) in the whole space,
one can use the Krein-Friedel formula \cite{F}.
The latter gives the contribution $\triangle N_\alpha(E)$
 of the scattering states to the change of the IDOS 
induced by the presence
of a scatterer, directly as the sum over phase shifts,
\begin{equation}
\triangle N_\alpha(E) =\frac{1}{\pi}\sum_l \delta_l(E)=
(2\pi i)^{-1}\ln\det\mbox{S},
\label{krein}
\end{equation}
S being the total on-shell S matrix.
As has been  shown in Sec.\ \ref{sec:cros}, the S matrix 
(\ref{sconv}) is singular as a consequence of
the singularity of phase shifts (\ref{shift}) which
in general do not decay when $E\rightarrow \infty$.
Therefore, the S matrix (\ref{sconv}) or (\ref{sconvm})
 cannot be substituted directly into
the Krein-Friedel formula (\ref{krein}). This can also be seen from
the fact that the sum (\ref{krein}) over phase shifts is not absolutely
convergent. We remind the reader that the sum  
of a series that is not absolutely convergent is ambiguous 
in a conventional sense. If suitably rearranged, the (conventional)
sum  of such a series can take any prescribed value.
Thus, to deal with such a series,  care must be taken.
In the present case of the AB potential we have found that
it is the $\zeta$-function regularization which 
gives the correct answer [(\ref{denreg})].
In the absence of bound states 
\begin{eqnarray}
\lefteqn{\ln\det\mbox{S}=\sum_{l=-\infty}^\infty 2i\delta_l
= i\pi\sum_{l=-\infty}^\infty (|l|-|l+\alpha|)}\nonumber\\
&& =
i\pi\left[ 
2\sum_{l=1}^\infty l -\sum_{l=-n}^\infty (l+\alpha)
+\sum_{-\infty}^{-n-1} (l+\alpha)\right]
\nonumber  \\
&& =
i\pi\left.\left[ 
2\sum_{l=1}^\infty l^{-s}-\sum_{l=0}^\infty (l+\eta)^{-s}
-\sum_{l=1}^\infty (l-\eta)^{-s}\right]\right|_{s=-1}
\nonumber  \\
&&
 =i\pi\left.
\left[2\zeta_R(s)-\zeta_H(s,\eta)-\zeta_H(s,1-\eta)\right]
\right|_{s=-1}= -i\pi\eta(1-\eta),
\label{frsum}
\end{eqnarray}
where $\zeta_R$ and $\zeta_H$ are the Riemann and the Hurwitz 
$\zeta$ functions (see Appendix \ref{ap:zeta}).
Thus, by using  Eq. (\ref{krein}),  the change of the DOS is
\begin{equation}
\triangle\rho_\alpha(E)=- \eta(1-\eta)\,\delta(E)/2.\nonumber
\end{equation}
This result is exactly (\ref{denreg}).
Therefore, despite the AB potential being long-ranged, 
the Krein-Friedel formula can  still be used when regularized
with the $\zeta$-function. In the case where  $\alpha$ is
 negative and equals $-n-\eta$,
the calculation of the change of the DOS is essentially the same.
The only change is that now
\begin{equation}
|l+\alpha|= \left\{
\begin{array}{rl}
l+\alpha, & l\geq n+1 \\
-l-\alpha, & l\leq n.
\end{array}\right.
\end{equation}
By using the Krein-Friedel
formula one again reproduces the result (\ref{denregs}).

As has been discussed in Sec.\ \ref{sec:self},
under the influence of bound states  
the phase shifts are changed only in two channels,
$l\!=\!-n$ and $l\!=\!-n-1$. Therefore, the contribution
of scattering states 
to the change of the IDOS is then
\begin{eqnarray}
\triangle N_\alpha(E)= 
&-&\frac{1}{2}\eta(1-\eta)+\frac{1}{\pi}\arctan
\left(\frac{\sin(\eta\pi)}{\cos(\eta\pi)-(|E_{-n}|/E)^{\eta}}
\right)
\nonumber\\
&-&\frac{1}{\pi}\arctan
\left(\frac{\sin(\eta\pi)}{\cos(\eta\pi)+
(|E_{-n-1}|/E)^{(1-\eta)}}\right),
\label{intsing}
\end{eqnarray}
where $E_{-n}$ and $E_{-n-1}$ are the binding energies 
in $l\!=\!-n$ and $l\!=\!-n-1$ channels. By using the arguments
given in the proof of relation (\ref{smtinv}) one finds that
the presence of bound states does not spoil the property
(\ref{denregs}) of the DOS, and  the IDOS is still a symmetric
function of $\alpha$,
\begin{equation}
\triangle N_{-|\alpha|}(E)=  \triangle N_{|\alpha|} (E). 
\label{intsings}
\end{equation}

Note that for 
$0<\eta<1/2$ the {\em resonance} appears at
\begin{equation}
E_{res}=\frac{|E_{-n}|}{\left[\cos(\eta\pi)\right]^{1/\eta}}>0.
\label{reso}
\end{equation}
The phase shift 
$\delta_{-n}(E)$ (\ref{shift}) changes by $\pi$ in the direction
of increasing energy and  the integrated density of states 
(\ref{intsing}) has a sharp
increase by $1$. The  profile of the resonance 
[the argument of arctan in  Eq. (\ref{shift})] is given by
\begin{equation}
 \frac{E^\eta \tan \eta \pi}{E^\eta - E_{res}^\eta}=
\frac{\Gamma}{E^\eta - E_{res}^\eta},
\label{resshape}
\end{equation}
where $\Gamma$ is the width of the resonance,
\begin{equation}
\Gamma = E_{res}^\eta \tan \eta \pi.
\label{reswidth}
\end{equation} 
Note that the profile (\ref{resshape})
 {\em is not} of the Breit-Wigner form,
\begin{equation}
\frac{\Gamma}{E-E_{res}}
\label{bfform}
\end{equation}
(see Ref.\  \cite{La}, \S 145).
\begin{figure}
\centerline{\epsfxsize=11cm \epsfbox{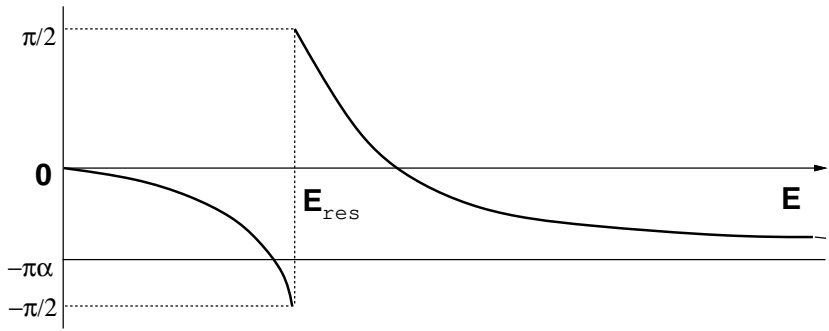}}
\caption{Typical energy dependence of 
$\triangle_l(E)$
in the channel with a bound state.}
\label{fig2}
\end{figure}
For $1/2<\eta<1$ the resonance is shifted to the 
$l\!=\!-n-1$ channel.
$\eta=1/2$ is a special point since resonances occur 
in both channels at infinity. Therefore,
the contribution
of the arctan terms in (\ref{intsing}) 
does not vanish as $E\rightarrow\infty$, but instead gives the value
$-1$. Here, we have assumed that bound states are in both the 
$l\!=\!-n$ and
$l\!=\!-n-1$ channels. As we shall see in Sec.\ \ref{sec:reg}, 
it is generally easier to form the bound state in the $l\!=\!-n$
channel than in the $l\!=\!-n-1$ channel. Therefore, unless $\eta$
is sufficiently large, only the bound state in $l\!=\!-n$ is present
and the above considerations have to be appropriately modified.
The above discussion also shows that even in the generic case,
when bound states are present, the DOS still depends only on the
distance from the nearest integer.
Equation (\ref{reso}) shows that if the energy of the 
bound state goes to zero, $E_b\uparrow 0$, the resonance also approaches
zero energy, $E_{res}\downarrow 0$. Therefore, in the limit
$E_b\uparrow 0$ the resonance merges with the bound state.
As will be discussed in more detail in Sec.\ \ref{sec:ren}
both the resonance and a bound state disappear from the point
 spectrum and leave behind the {\em phase-shift flip}.

Having calculated phase shifts (\ref{shift}) for a generic self-adjoint extension, one can calculate the time delay $\triangle t$ associated
with the energy derivative of the phase shift \cite{RN},
\begin{equation}
\triangle t= 2\hbar\frac{d\delta(E)}{dE}\cdot
\label{tdelay}
\end{equation}
One sees immediately that if $l$ is different from $-n$ and $-n-1$,
then the time delay is $\triangle t=0$. 
The time delay (\ref{tdelay}) can be nonzero
only in the channels $l=-n$ and $l=-n-1$, and at the 
energy corresponding to the resonance the time delay is infinite.

\section{Regularization}
\label{sec:reg}
To identify the physics that underlines different 
self-adjoint extensions
we have considered the situation when the AB flux tube
is regularized by a flux tube of a finite radius $R$, 
and the magnetic field $B$ inside it satisfies 
the constraint (\ref{bconst}). This situation was discussed
first in Ref.\ \cite{Kre}. The discussion of this case
has a direct relevance for experiment since vortices
usually realized in experiments are not singular and do have
a nonzero radius \cite{BKP,WW,RS}.
In order for a  bound state to exist, the matching equation for
logarithmic derivatives of the exterior (see Appendix \ref{asymp})
and interior (see Appendix \ref{apehyper}) solutions 
in the $l$th channel must have a solution.
The logarithmic derivative of the exterior solution (\ref{bounst})
is given by 
\begin{equation}
O_l(x) = x\frac{K_{|l+\alpha|}'(x)}{K_{|l+\alpha|}(x)}\leq -|l+\alpha|.
\label{logex}
\end{equation}
It depends only on the flux $\alpha$ and not on a 
particular regularization of the interior of the flux tube.
Here, parameter $x$ is given by
\begin{equation}
x=\kappa R= \frac{\sqrt{2m|E_b|}}{\hbar}\,R,
\label{xparam}
\end{equation}
where $E_b$ is the bound state energy.
$O_l(x)$ {\em decreases} from $-|l+\alpha|$ to $-\infty$ as 
$x\rightarrow\infty$ (see Appendix \ref{asymp}). Therefore,
\begin{equation}
O_l(x)\leq \left\{
\begin{array}{rl}
-\alpha+|l|,&l\geq -n\\
\alpha-|l|,&l\leq -n-1.
\end{array}\right.
\label{oineq}
\end{equation}

In contrast, the logarithmic derivative of the
interior solution depends on both $\alpha$ and the particular
distribution of the magnetic field inside the flux tube.
For example, in the case of a {\em homogeneous} 
field regularization (see, e.\ g., Ref.\ \cite{BV}) 
and in the absence of the magnetic moment coupling
($g_m=0$) or any other additional interaction inside the flux tube
the logarithmic derivative of the 
interior solution (\ref{psiint}) is 
\begin{equation}
I_l(x)= -\alpha+|l|+2 \alpha\,\frac{a}{b}\,
\frac{M\left(a+1,b+1,\alpha\right)}
{M\left(a,b,\alpha\right)},
\label{lin}
\end{equation}    
where $M(a,b,\alpha)$ is the Kummer hypergeometric 
function \cite{AS}, and 
\begin{eqnarray}
a &=& \frac{|l|+l+1}{2}+(x^2/4\alpha)\geq \frac{1}{2},\nonumber\\
b &=& |l|+1\geq1.
\label{abdef}
\end{eqnarray}
For those values of $a$ and $b$ one obtains, by using Eq.\ (\ref{mrel}), 
\begin{equation}
\frac{2a}{b}\, \frac{M(a+1,b+1,\alpha)}{M(a,b,\alpha)}>0.
\label{mmrel}
\end{equation}
Hence, 
\begin{equation}
I_l(x)>-\alpha+|l|\geq -|l+\alpha|
\label{iineq}
\end{equation} 
and it is impossible to get a bound state in this case
because the matching equation,
\begin{equation}
O_l(x)= I_l (x),
\label{matching}
\end{equation}
cannot be satisfied for any $x$. 
Since
\begin{equation}
M(0,b,\alpha)=1,
\label{0mcon}
\end{equation}
one finds that Eq. (\ref{matching}) might have  
a solution provided $a=0$ and $x=0$.
Now, our purpose will be to break the constraint
(\ref{abdef}) on $a$ which implies (\ref{mmrel}). In particular, 
one can show that parameter $a$ can become {\em negative}
if an {\em attractive} potential $V(r)$ 
is placed inside the flux tube,
\begin{equation}
V(r)|_{r\leq R}=-\frac{\hbar^2}{2m}\frac{\alpha}{R^2}c(R),
\label{tubpot}
\end{equation}
and $V(r)=0$ otherwise. This amounts to changing $x^2/4\alpha$ 
in (\ref{abdef}) to $x^2/4\alpha -c/4$. 
If one writes $c(R)$ as $c(R)=2[1+\varepsilon(R)]$ one finds 
\begin{eqnarray}
a &=& \frac{|l|+l}{2}+\frac{x^2}{4\alpha} - 
\frac{\varepsilon(R)}{2},\nonumber\\
b &=& |l|+1.
\label{abedef}
\end{eqnarray}
The attractive potential can be either put in
by hand, or, if the Pauli Hamiltonian (\ref{pauli})
is used, as arising
from the magnetic moment coupling of the electrons 
with the spin  opposite to the direction of magnetic field $B$.
In the latter case $\varepsilon(r)=$ const is determined
by the anomalous part of the magnetic moment,
\begin{equation}
\varepsilon=(g_m-2)/2.
\end{equation}
Equation (\ref{abedef}) shows that the critical value of $a=0$ 
corresponds to the case when $g_m =2$, i.\ e., the case when the
magnetic moment is not anomalous, and $l\leq 0$.
In this case the matching equation (\ref{matching}) 
does have a solution at $x=0$ for $0\geq l\geq -n$.
If one substitutes the value of $x$ into
either the exterior (\ref{bounst}) or the interior (\ref{psiint})
solution one would obtain a nonsense: ``solutions'' that 
do not depend on $r$ (in the case of the exterior solution 
the limit $x\rightarrow 0$ is in fact
singular). A more subtle method is needed \cite{AC} to show
that in this case zero modes do appear. 
Since the magnetic field is not singular anymore
the Aharonov-Casher theorem \cite{AC,Th} applies.
The Aharonov-Casher theorem \cite{AC} tells us
that in a general finite-flux magnetic field $B({\bf r})$,
for which
\begin{equation}
 \int_\Omega B({\bf r})\, d^2 {\bf r} = \Phi = \mbox{const},
\label{bconst}
\end{equation}
the $2D$ Pauli Hamiltonian  at $g_m=2$ has
exactly $]\alpha[-1$  {\em zero modes},
\begin{equation}
e^{-\phi(r)},\, e^{-\phi(r)}(x_1 -ix_2),\,\ldots,\, 
e^{-\phi(r)}(x_1 -ix_2)^{n-1}.
\label{zerom}
\end{equation}
Here, $]\alpha[$ stands for the nearest integer 
{\em larger} than or equal to $\alpha$, and, 
for a given magnetic field $B({\bf r})$, the function $\phi(r)$
is defined by
\begin{equation}
\phi({\bf r})=\frac{e}{h c}\int_{R^2}\ln|{\bf r}-{\bf r}'|\,
B({\bf r}') \,d^2{\bf r}' .
\label{phidef}
\end{equation}
If $\alpha$ is an integer $n$, then the number of zero modes is $n-1$;
if not, their number is $n=[\alpha]$. For $\alpha\leq 1$ no 
zero mode is present in the spectrum \cite{AC,Th}.
%
The proof of the theorem uses the fact that the Pauli Hamiltonian 
for $g_m=2$ can be written as the square of the 
Euclidean two-dimensional massless Dirac operator \cite{AC}.
The result only depends on the total flux $\alpha$ and
not on a particular distribution of a magnetic field $B$.
The source of the attractive potential 
(\ref{tubpot}) inside the flux
tube is not important to the formation of zero modes.

Now, if $\varepsilon>0$, i.\ e., $g_m>2$,
 the parameter $a$ can even become negative.
One has $a<0$ for $l\leq 0$, $\varepsilon> 0$, and 
$x\in \left[0,\sqrt{2\alpha\varepsilon}\,\right)$.
We shall show that there are at least $n+1$ bound states for any
finite $R$ in this case.
In other words, if the attractive potential
(\ref{tubpot}) is situated inside the flux tube, 
the coupling with the interior of the flux tube
 becomes sufficiently strong for the particle to be confined
(on the cyclotron orbit) {\em inside} it:
the wave function (\ref{bounst}) of the bound state 
decays exponentially outside the flux tube.
In what follows we shall confine ourselves to $l\leq 0$.
The reason is that in order that $a$ be negative for $l=1$,
we must choose $g_m=2+2\varepsilon>6$.   
This is the value of $g_m$ that is out of experimental interest.
Obviously, if $l>1$, then the minimal value of $g_m$ that does $a$  negative
is proportional to $l$ and hence larger than the minimal value of $g_m$
for $l=1$. The number of bound states is given by the number of channels
in which the matching equation (\ref{matching}) can be satisfied
with a solution $x_l>0$. The matching equation (\ref{matching})
implies that the ratio in Eq.\ (\ref{mmrel}) 
has to be {\em negative},
\begin{equation}
\frac{a}{b} \frac{M(a+1,b+1,\alpha)}{M(a,b,\alpha)}<
\left\{
\begin{array}{ll}
\ 0,&l\geq -n\\
-|l|+\alpha,&l\leq -n-1.
\end{array}\right.
\label{mmrelm}
\end{equation}
This can be satisfied for $0\geq l\geq -n$ and $-1<a<0$ 
only if (see Appendix \ref{apehyper})
\begin{equation}
M(|a|,b,\alpha)<2.
\label{excon}
\end{equation}
The latter constraint on the values of $a$ does not impose
any physical restrictions since it allows for $g_m\in[2,6)$, 
i.\ e., for almost all realistic values of $g_m$.
Since Eq. (\ref{0mcon}) holds,
it is clear from Eq.\ (\ref{abedef}) that Eq.\ (\ref{excon}) can 
always be satisfied. It is sufficient to look for  
a bound state energy
$E_b$  having roughly the form [see  Eqs. (\ref{xparam}) 
and (\ref{abedef})]
\begin{equation}
|E_b|= 2\alpha\, \frac{2m}{\hbar^2 R^2}\, (\varepsilon - t),
\label{boundep}
\end{equation}
where $t$ ($0\!<\!t\!<\!<\!\varepsilon$) is some small positive number.
Therefore, one concludes that there are at least $n+1$ bound states
at any finite $R$ (cf. Refs.\ \cite{BV,BV1}). 
To satisfy Eq.\ (\ref{mmrelm}) for $l\leq -n-1$ is more difficult
because the condition becomes more restrictive.
Indeed, it is sometimes stated in the literature that for $n=0$
and homogeneous regularization only one bound state can exist
(see, e.\ g., Refs.\ \cite{AG} and \cite{BV}). We shall show that although
generally it is true, in some circumstances the $(n+2)$th
bound state in the $l\!=\!-n-1$ channel 
{\em does} appear. It is clear from Eq.\ 
(\ref{mmrelm}) that the condition to be satisfied gets weaker
as $1-\eta$ gets smaller. In particular, one always can choose
$x$ such that for $\eta\in(1-\sigma,1]$, $0<\sigma\!<\!<\!1$,
\begin{equation}
M(|a|,b,\alpha)\leq 2-\gamma
\end{equation}
[see Eq.\ (\ref{0mcon})], where $0<\gamma\!<\!<\!1$ is some small number. 
Then  for these values of $x$ and $\eta$, the left-hand side of
Eq.\ (\ref{mmrelm}) is uniformly bounded,
\begin{equation}
\frac{a}{b} \frac{M(a+1,b+1,\alpha)}{M(a,b,\alpha)}< 
-\tilde{\gamma}<0.
\end{equation}
Since $\alpha-|l| \rightarrow 0$ when $\eta\rightarrow 1$, 
there exists a 
$\tilde{\sigma}$, $0<\tilde{\sigma}\leq\sigma$,
such that Eq.\ (\ref{mmrelm}) is satisfied for a given $x$
and $\eta\in (1-\tilde{\sigma},1)$ in the $l\!=\!-n-1$.
Therefore, the actual number
of bound states can in principle be higher than $n+1$  since
the condition (\ref{mmrelm}) can be satisfied even for some 
$l\leq -n-1$. An illustrative example is provided by a 
{\em cylindrical shell} regularization \cite{Hag,BV} 
in which the magnetic field is given by
\begin{equation}
B(r)=\frac{\alpha}{2\pi R}\,\delta(r-R).
\label{cylshell}
\end{equation}
As has already been discussed, the exterior solution (\ref{bounst})
is the same as in the homogeneous field
regularization. Only the interior solution changes. We shall 
denote its logarithmic derivative by the superscript `$c$'.
By using the results of Ref.\ \cite{BV} it can be shown 
that the matching equation takes  the form
\begin{equation}
O_l(x)=I^c_l(x)=-\alpha +|l|- \alpha\,
\varepsilon+\frac{1}{2b} x^2 +{\cal O}(x^4),
\label{matcho}
\end{equation}
where $b$ is as defined by Eq.\ (\ref{abdef}) or 
Eq.\ (\ref{abedef}). As in the previous case, the matching 
equation (\ref{matcho}) for $g_m>2$ can be satisfied for
 $0\geq l\geq -n$. 
However, apart from these values of $l$, Eq.\ (\ref{matcho}) 
can be satisfied for $l$ below $l\!=\!-n$ provided that
\begin{equation}
|l|-\alpha<\frac{\alpha\,\varepsilon}{2}\cdot
\label{num1}
\end{equation}
Note again that when  $l\!=\!-n-1$
one does not generally have  a bound state, as is also the case 
of the cylindrical shell regularization. However, if
\begin{equation}
\eta>\frac{1-n\varepsilon/2}{1+\varepsilon/2},
\end{equation}
a bound state does appear in the $l\!=\!-n-1$ channel.
As is seen from Eq. (\ref{num1}), when $n$ increases, this bound state
appears for ever smaller $\eta$. Eventually, for
\begin{equation}
n\geq n_0=]2/\varepsilon[
\end{equation}
the restriction (\ref{num1}) on $\eta$ disappears. 
Moreover in this case bound states can appear, even 
for a positive $l$,  provided that
\begin{equation}
l<\frac{\alpha\,\varepsilon}{2},
\label{num2}
\end{equation}
or, equivalently, that
\begin{equation}
\alpha\geq \frac{4}{g_m-2}\cdot
\end{equation}
However, by taking into 
account that for the electron $g_m-2=0.002\ 32$, the flux has to be
of order $\sim 2000$ for this to be the case. 
To conclude, according to Eqs.\ (\ref{num1}) and (\ref{num2})
the number of bound state 
$\#_b$ in the cylindrical shell regularization is
\begin{equation}
\#_b =1 +n + \left[\alpha(g_m-2)/4\right] +
\left[\alpha(g_m+2)/4 -n\right].
\label{numberb}
\end{equation}                     
Here $[.]$ denotes the integer part.  The number (\ref{numberb})
of bound states in the cylindrical shell regularization 
is generally {\em higher} than that in the homogeneous 
field regularization. 
One can understand the physical origin of this difference in a 
simple way. In the cylindrical shell regularization, the energy
$E_B$ of magnetic field is {\em infinite} for any $R\neq 0$ 
and in this sense the magnetic field inside the flux tube 
is much stronger than, for example,
in the homogeneous field regularization when
\begin{equation}
E_B=\frac{\pi}{2} B^2 R^2 = \frac{\Phi^2}{2\pi R^2}
\label{mgen}
\end{equation}
stays {\em finite} for any nonzero $R$.
Therefore, in contrast to the number of zero modes 
given by the Aharonov-Casher theorem \cite{AC,Th}, 
the number of bound states {\em does depend} not only on the total
flux $\alpha$ but on a particular distribution of the magnetic field $B$,
and hence on the energy of the magnetic field, also.
A similar check with the $1/r$ regularization (see Ref.\ 
\cite{BV}) allows
us to make the hypothesis that their number is less than or 
equals to $\#_b$ and that the bound is saturated when one uses  
the cylindrical shell regularization of the AB potential. 

In two dimensions for $R\neq 0$ and $\lambda$ arbitrarily
small,  the Schr\"{o}dinger equation  always has a bound state 
in the potential $\lambda V(r)$ (\ref{tubpot}) 
in the absence of the AB potential \cite{La}. For $R\neq 0$ the 
potential $\lambda V(r)$  is not singular and the wave function
of the bound state is not singular, either. Now, if
the AB potential is put on top of $\lambda V(r)$ 
the bound state in general {\em disappears} in its presence.
Therefore, the discussion in this section implies
that, generally, the AB potential has a {\em deconfining} effect
on the bound state \cite{AG}. Provided $\lambda\geq 2$
and in the presence of the AB potential the
wave function of a bound state (\ref{bounst})
is {\em singular} and the phase shifts [see  Eq. (\ref{shift})]
are changed.  Then, in the limit $\eta\rightarrow 0$, the singular 
wave function (\ref{bounst})
becomes the regular one and the phase shifts 
must have their conventional values (\ref{convshift}). Only 
the singular bound state wave function 
necessitates the change of phase shifts. This explains why
$A_l\rightarrow 0$ in (\ref{psising}) as $\eta\rightarrow 0$, 
which has already been used above in the calculation of
$\triangle_l$.

\section{The $R\rightarrow 0$ limit and the
interpretation of self-adjoint extensions}
\label{sec:ren}

In this section we shall examine the limit $R\rightarrow 0$
subject to the condition (\ref{bconst}).
In the case when the flux tube is exterior to the 
system and particles are not allowed to interact with 
its interior, the $R\rightarrow 0$ limit
is trivial as there are neither zero modes nor bound states.
Therefore, in what follows we shall confine ourselves to the
case when the flux tube is a part of the system and particles
do interact with its interior. In the rigorous mathemetical 
sense the limit $R\rightarrow 0$ 
is described by some self-adjoint extension, and we shall 
discuss a correspondence between the $R\neq 0$ case and the limiting case.

In the limit $R\rightarrow 0$ the potential $V(r)$ as defined 
by Eq. (\ref{tubpot}) goes formally to the $\delta$ function,
\begin{equation}
V(r)|_{r\leq R} \rightarrow -[1+\varepsilon(0)]\,
\frac{\hbar^2}{m}\frac{\alpha}{r}\,\delta(r).
\label{tubpol}
\end{equation}
As has been shown above, until the
magnetic moment $g_m$ reaches its critical value $g_m=2$, nothing
is changed with respect to the case of the impenetrable flux tube,
and the limit $R\rightarrow 0$ is again trivial. The limit becomes
nontrivial at the critical coupling when $]\alpha[-1$ zero modes
(\ref{zerom}) exist \cite{AC,Th}. 
In this case the symmetry of the spectrum 
with regard to $\alpha\rightarrow\alpha\pm 1$ is lost. 
Then, as $R\rightarrow 0$ we shall show that zero modes (\ref{zerom})
disappear from the point spectrum and merge with the continuous 
spectrum. Indeed, in the limit $R\rightarrow 0$ the magnetic 
field $B({\bf r})$ is given by
\begin{equation}
B({\bf r}) =2\pi \alpha \,\delta({\bf r})=\frac{\alpha}{r}\,\delta(r).
\end{equation}
In this case, $\phi({\bf r})$ defined by 
Eq.\ (\ref{phidef}) can be calculated  exactly. One finds
\begin{equation}
\phi({\bf r})=\alpha\ln|{\bf r}|.
\end{equation}
The zero modes (\ref{zerom})
are then obviously singular 
at the origin and they  are not elements
of $L^2[(0,\infty), rdr]$. As $R\rightarrow 0$,  
zero modes (\ref{zerom}) get more and more singular at 
the position of the flux tube and
eventually, at the limit, they become nonintegrable and merge
with the continuous spectrum. In the latter
case one has to check for square integrability not only at infinity
but at the position of singularities of the field, too.
It is here where the theorems fail. 
Another argument for the disappearing of zero modes at the $R\rightarrow 0$ 
limit is to note that the only two channels at which the spectrum can 
differ from that of an impenetrable flux tube are the channels with
$l=-n$ and $l=-n-1$.
However, for any $R\neq 0$, zero modes never occur in these
two channels. Instead, they are in channels with $0\geq l\geq -n+1$
\cite{AC,Th}.
If one suspects that zero modes can appear in some  channels different
from those given in Eq.\ (\ref{zerom}), one can check directly
that for whatever $l$, the functions given by Eq.\ (\ref{zerom}) are
not square integrable either at infinity or at the origin.
Therefore, in the limit $R\rightarrow 0$ the symmetry of the spectrum 
under the substitution  
$\alpha\rightarrow\alpha\pm 1$ is again
recovered. The best method of illustrating this point
is to consider a situation when $g_m-2=2\varepsilon>0$ and stays
constant as $R\rightarrow 0$. 
Whenever $g_m>2$ [or $g_m=2$ with an attractive potential 
$V(r)=-\varepsilon/R^2$, $\varepsilon>0$ arbitrary small]
bound states occur in the spectrum.
They correspond to solutions $x_l>0$ of  Eq. (\ref{matching}).
Note that the solutions are only a function of $\alpha$. They
do not change when $R$ changes. Since $x$ is given by 
Eq. (\ref{xparam}) one finds that as $R\rightarrow 0$ the bound state
energy $E_b$ scales as
\begin{equation}
E_l=-\frac{\hbar^2 x_l^2}{2mR^2}\cdot
\label{bsen}
\end{equation}
In other words, in addition to the breaking the symmetry
$\alpha\rightarrow\alpha\pm 1$ of the spectrum, in the presence of bound
states the scale invariance is also broken \cite{MT,SI}. 
 Nevertheless, one finds that bound states {\em decouple}
in the $R\rightarrow 0$ limit from the Hilbert
space $L^2[(0,\infty), rdr]$ and take away the 
{\em nonperiodicity} of the spectrum
(under $\alpha\rightarrow\alpha\pm1$) that
persists for any finite $R$. What is left behind is nothing but the 
{\em conventional} AB problem with the change of the density
of states given by (\ref{denreg}).  To show this, note 
that $A_l\rightarrow 0$ in the general scattering
solution (\ref{psichan1}) in 
the limit $\kappa\rightarrow\infty$. Hence, in the limit
the regular state (\ref{regpsi}) is recovered. Another argument
is to note that the bound state (\ref{bounst}) is a 
function of $\kappa r$
and decays exponentially as $\kappa r\rightarrow \infty$. Therefore,
since $\kappa\rightarrow \infty$ as $R\rightarrow 0$, 
the wave function goes  to zero and thereby disappears 
from the spectrum. The electron is confined inside a flux
as in a black hole and ceases to communicate with an outside
world. 

The above fact might at first be surprising, however
it has been demonstrated by Berezin and
Faddeev \cite{BF} more than $30$ years ago that a 
nontrivial limit $R\rightarrow 0$ for a $\delta$-function potential
exists only if the coupling constant is {\em renormalized}.
The latter is necessary for a proper mathematical definition
of the Schr\"{o}dinger operator 
with the $\delta$-function potential \cite{BF,AGHH}.
In other words, in order to obtain the bound state 
in the limit $R\rightarrow 0$
it is necessary that 
\begin{equation}
\varepsilon(R)\rightarrow 0
\end{equation} 
as $R\rightarrow 0$ in Eq. (\ref{tubpot}) \cite{BV,MT,Sve}. 
The actual energy of the bound state then depends 
on the details of the interaction and the details 
of renormalization.
Obviously, in this case the symmetry of the spectrum
under $\alpha\rightarrow\alpha\pm 1$ is broken.
Since the spectrum and phase shifts do not change until
the attractive potential inside the flux tube is renormalized
to its critical strength and a bound state is formed, this
generalizes the result of Ref.\ \cite{AU} that the AB 
scattering in the case of open boundary conditions 
at the flux tube boundary coincides with the AB scattering with
Dirichlet  boundary conditions.
In other words, provided the attractive potential inside 
the flux tube   is not renormalized
to its critical value and remains either weaker or stronger 
than the critical potential, then the flux tube remains impenetrable
in the limit $R\rightarrow 0$.                     

If a bound state is present, the phase shift in a corresponding
channel acquires an energy dependent term (\ref{trianl}).
The latter, in the limit of zero bound energy, gives rise
to the phase-shift flip.
Therefore, it is natural to assume that the change
(\ref{trianl}) of the phase shift or the phase-shift
flip will occur in the limit $R\rightarrow 0$ only in these channels
where the bound state occurs for $R\neq 0$.
When the phase-shift flip takes place, the symmetry
$\alpha\rightarrow\alpha\pm 1$ is again broken.
Our calculation  [see relations (\ref{shift}) and
(\ref{shflip})] shows clearly that the phase-shift flip \cite{Hag} is 
not connected to the spin but may occur in its absence 
as well. 

As has been mentioned, solutions of the matching equations
are only functions of the flux, $\alpha$ and $x(=\kappa R)$.
On the other hand, scattering solutions are functions
of the flux $\alpha$, the ratio $k/\kappa$
[see Eq.\ (\ref{al})], and $kr$. From the experimental point 
of view it is useful to remark the following {\em duality}:
the physics at a given radius $R_1$ of the flux tube and at
momentum $k_1$ is identical to that at $R_2$ and $k_2$,
provided
\begin{equation}
k_2=\frac{k_1 R_1}{R_2}\cdot
\label{dual}
\end{equation}
When Eq.\ (\ref{dual}) holds, then the relative combination of
the regular and the singular Bessel functions in Eq.\ 
(\ref{psising}), and hence the phase shift, are the same.
Therefore, provided one has only a vortex of a finite radius $R_1$ at 
disposal, one can examine the physics of almost singular
vortices with a radius $R_2\!<\!\!<\!R_1$
by performing experiments at very {\em small momenta} $k$,
i.\ e., such that $R_1\!<\!\!<\! 1/k$.
It is in the latter situation where the phase-shift flip has
been established \cite{MRW}.  Moreover, it is easier
to realize experimentally.

\section{Energy calculations}
\label{sec:en}
In this section we shall assume that $g_m$ is a fixed constant and 
that no
renormalization of $g_m$ occurs. 
The energy  of the system consisting of particles and
field will be evaluated for a sequence of flux tubes of decreasing
radii subject to the constraint that the total flux $\Phi$ (\ref{bconst})
is the same in each.
Therefore,  dynamical phenomena such as the induction of the electric field
and return fluxes will be ignored. One reason for this rough
approximation is that we do not know better.

 Let us now discuss the  cases where $g_m$ is, respectively, less than, 
equal to, or greater than 2.
It has been already shown that
up to $g_m=2$ no bound state is present in the spectrum
and the change of the density of the
scattering states is still given by  Eq. (\ref{denreg}). 
Zero modes which occur for $g_m=2$ at $R\neq 0$ are regular at 
the origin and do not
change phase shifts as $R\rightarrow 0$.
Because the energy of the magnetic field (\ref{mgen}) 
tends to infinity as $R\rightarrow 0$
the system consisting of particles and
field is definitely {\em stable} 
with respect to spontaneous creation of the AB field.

When $g_m>2$, then bound states occur in the spectrum.
Their energy is given by  Eq. (\ref{bsen})
and scales to $-\infty$ when $R\rightarrow 0$,
in the same way (as $1/R^2$) that the magnetic field energy (\ref{mgen}) 
does to $+\infty$.
As has been shown in Sec. \ref{sec:ren},
provided $\varepsilon(R)$ is not renormalized in the limit
$R\rightarrow 0$, the bound states decouple in 
the limit from the Hilbert
space $L^2[(0,\infty), rdr]$ [$A_l\rightarrow 0$ in  Eq. 
(\ref{psising}] in 
the limit) and take away the {\em nonperiodicity} of the spectrum
with regard to $\alpha\rightarrow\alpha\pm1$ that
persists for any finite $R$. 
What is left behind is nothing but the 
{\em conventional} AB problem with a change of the density
of states (\ref{denreg}).  

%
In what follows the homogeneous field regularization will be used.
Note that the homogeneous magnetic field optimizes the energy functional
\begin{equation}
E=\int_\Omega B^2({\bf r})\,d^2 {\bf r}
\end{equation}
subject to the constraint (\ref{bconst}).
Bound state solutions $x_l$ for the homogeneous
field regularization  determine the function $X(\alpha,g_m)$,
\begin{equation}
X(\alpha, g_m)=(4\pi\alpha^2)^{-1}\sum_l x^2_l(\alpha,g_m)\geq 0.
\end{equation}
By comparing the coefficients in front of $1/R^2$ in 
Eqs. (\ref{mgen}) and (\ref{bsen}), one finds that whenever
 the ratio of the rest energy to the electromagnetic
energy is less than $X(\alpha,g_m)$,
\begin{equation}
\frac{mc^2}{e^2} <X(\alpha,g_m),
\label{instab}
\end{equation}
the total energy of field and matter together goes to $-\infty$
as $R\rightarrow 0$.
Therefore, in the {\em static} approximation, without the account of
the electric field energy, the energy of the system 
decreases with decreasing $R$.
This does not show the instability against the 
spontaneous creation of a magnetic field yet, since the full treatment
has to take  the dynamics and the magnetic moment form factors
into account. Nevertheless, the discussion shows that the case
of $g_m>2$ is different with respect to $g_m\leq 2$.
In three space dimensions,  the function $X(\alpha,g_m)$ is replaced
by $X(\alpha,g_m)/L$, with $L$ the length of the flux string,
and one has to discuss the density of states for a long flux ring
\cite{OP}.

%
%
Note, in passing, that in the relativistic quantum-mechanical 
treatment \cite{AM89}
one finds that the system is stable.
The reason is as follows:
for spin-up  electrons the magnetic moment coupling
introduces a repulsive interaction and, hence, 
there is no natural way to obtain
a spectrum with bound states as there was 
 for spin-down electrons in which case
this interaction is attractive.
To obtain a bound state in the spectrum,  an attractive potential
$2V(r)$ has to be put inside the flux tube by hand.
This is the principal reason why in the case of the Dirac electron
(a ``pair" of the spin-up and spin-down Schr\"{o}dinger electrons)
the magnetic moment coupling cannot produce a bound state
in the spectrum no matter how large or small the magnetic
moment is \cite{AM89}. However, there is one exception and
this occurs when the gauge field is the Chern-Simons field \cite{DJT}.
Indeed, it has recently been discussed by Hosotani \cite{YH} that
in the full-fledged quantum-field theory model with the
Chern-Simons gauge field \cite{DJT}
a magnetic field can be spontaneously generated.
The Chern-Simons field is somewhat pathological with respect to the 
discussion in this section, for in this case the density of matter acts as the source of 
the AB gauge field,
and all particles carry a flux \cite{DJT}.
In this respect one has  ``spontaneous magnetic field generation"
whenever the particle density is different from zero.
The discussion that led us to the stability condition
(\ref{instab}) remains reasonable even in this case.
Since  the energy of the Chern-Simons 
field is zero, there is nothing to impede the formation of a magnetic field.  Regarding massless charged particles, 
note that the result
of Gribov \cite{VG} concerning an  instability of 
massless charged particles
shows that the ratio of the rest energy to the electromagnetic
energy is an important parameter in field theory, and a
condition similar to (\ref{instab})  must hold.
The massless charged particles were claimed by Gribov not to
exist in nature, since  they are completely
screened locally in the process of their formation.

\section{The Hall effect in the dilute vortex limit}
\label{sec:hal}
\begin{figure}
\centerline{\epsfxsize=11cm \epsfbox{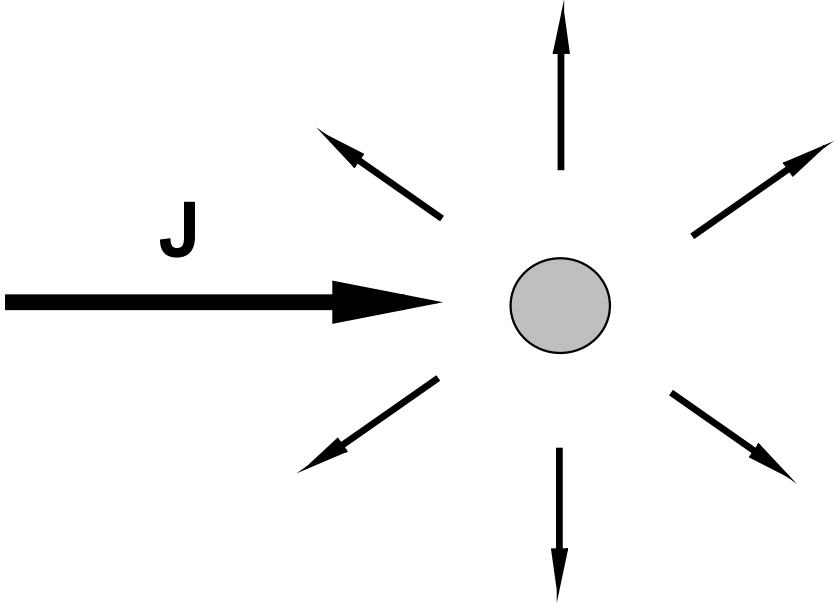}}
\caption{The differential scattering cross section $d\sigma(k,\varphi)$ gives 
the fraction of an incident current {\bf J} which is scattered
to the angle $\varphi$. In the case of an asymmetric differential 
scattering cross section there is generally a surplus of particles
flowing in one of the transverse directions.}
\label{fghal}
\end{figure}

As has been discussed in Sec.\ \ref{sec:cros}, the differential
scattering cross section (\ref{dbcross}) for a generic self-adjoint
extension is asymmetric with regard to 
$\varphi\rightarrow-\varphi$. One can show that the asymmetry can have 
important experimental
consequences: in contrast to the conventional symmetric
differential cross section (\ref{dcross}), the asymmetric
differential
cross section (\ref{dbcross}) can give rise to the Hall effect.
Indeed, if the incident wave function is normalized
to unit current density, the differential 
scattering cross section, $d\sigma(k,\varphi)$, is nothing
but the transition probability between the incident 
scattering state and the scattering state that propagates
in a direction $\varphi$ with respect to the incident wave
\cite{La}. In other words, the differential scattering cross section
gives the fraction
of particles from the incident beam that are scattered off
to the angle
$\varphi$. Now, let us consider electrons  propagating with
the Fermi momentum $k_F$ in a sample
in a direction singled out by an applied electric field. 
If vortices are  randomly distributed throughout the sample the 
electrons will be
scattered. In what follows, the {\em dilute vortex limit}
will be considered, 
in which the multiple-scattering contributions 
are neglected. Results concerning the Hall effect are then obtained
by summing over the single-vortex contributions.
The asymmetric differential
scattering cross section (\ref{dbcross}) of an electron from a vortex
means that, generally, there 
is a net surplus of the electrons propagating in one of the transverse directions, i.\ e., either to the right or to the left
(see Fig.\ \ref{fghal}).
The quantity that measures the fraction of the electrons moving
in a transverse direction is $\sin(\varphi)\, d\sigma(k_F,\varphi)$.
Therefore, if the density of vortices is $n_v$,
the Hall current, in the dilute vortex limit,
is proportional to
\begin{equation}
n_v\, \int_{-\pi}^\pi\!
d\varphi\, \sin\varphi\, \frac{d\sigma}{d\varphi}(k_F,\varphi).
\label{holik}
\end{equation}
By inverting the conductivity tensor one finds that 
the Hall resistivity, $\rho_{xy}$, is
\begin{equation}
\rho_{xy}=  \rho_H^0 \frac{k_F}{\alpha} \int_{-\pi}^\pi\!
\frac{d\varphi}{2\pi}\sin\varphi\, \frac{d\sigma}{d\varphi}(k_F,\varphi),
\label{hall}
\end{equation}
which was obtained by Nielsen and Hedegaard \cite{NH}.
In the latter case the result (\ref{hall})
was obtained by solving (in the dilute vortex limit)
Boltzmann's equation,
which relates the scattering and the transport properties. Here, 
$\rho_H^0=Bc/n_e e$ is the Hall resistivity in a uniform
magnetic field $B$, $n_e$ is the density of electrons, 
and $e$ is the electronic charge. The uniform magnetic field $B$ 
is obtained by averaging
over the field produced by vortices with the density $n_v$,
\begin{equation}
B=n_v \alpha\Phi_0.
\end{equation}
Fortunately, because of $\sin\varphi$ factor in  Eq. (\ref{hall}),
the differential scattering cross section is only needed 
for $\varphi\neq 0$ to determine the Hall resistivity.
When the differential scattering cross section
(\ref{dbcross}) is inserted in Eq. (\ref{hall}) one finds 
that only the last term contributes and 
\begin{equation}
\rho_{xy}= \frac{4n_v}{n_e}\frac{hc^2}{e^2}\sin(\pi\alpha)\left[
\sin\triangle_{-n} \cos(\triangle_{-n}-\pi\alpha)
+\sin\triangle_{-n-1} \cos(\triangle_{-n-1}+\pi\alpha)\right].
\label{reshall}
\end{equation}
Equation (\ref{reshall}) shows that one needs more than the asymmetry
of the differential scattering cross section for the Hall resistivity
to be different from zero. In fact, the Hall resistivity,
$\rho_{xy}$, vanishes whenever (modulo $\pi$)
\begin{equation}
\triangle_{-n}=-\triangle_{-n-1}.
\label{ineqd}
\end{equation}
However, as has been discussed in Secs. \ref{sec:reg} and
\ref{sec:ren}, in any realistic situation relation (\ref{ineqd})
is not generic and the Hall effect will appear.
For example, in the case where the phase-shift flip occurs only 
in the $l=-n$ channel, i.\ e., $\triangle_{-n}=\eta\pi$ and $\triangle_{-n-1}=0$, one finds
\begin{equation}
\rho_{xy}= \frac{4n_v}{n_e}\frac{hc^2}{e^2}\sin(\pi\alpha)
\sin (\eta\pi) \cos(\pi n).
\label{reshall*}
\end{equation}
In this case  it is easy to check  that if the vorticity $\alpha$ 
increases, $\rho_{xy}$ does not change its sign.
Our result (\ref{reshall}) shows that the Hall resistivity 
is proportional to the density of vortices and depends 
on their vorticity via trigonometrical functions. As a self-consistency 
check, the Hall resistivity (\ref{reshall}) vanishes
whenever  $\alpha$ is an integer. 
In the case of vortices in a type II superconductor
\cite{BKP}, if the magnetic field increases,
the vorticity of each vortex remains constant, and only their
density $n_v$ changes linearly with the applied field.
Therefore, the dependence of the Hall resistivity, $\rho_{xy}$,
on the magnetic field is  {\em linear} in the dilute vortex limit
\cite{RS}. Obviously, if the magnetic field is sufficiently large,
the dilute vortex limit ceases to be valid and deviations 
appear \cite{BKP}. Nevertheless, measurements of the Hall effect
on a single isolated vortex is now almost experimentally
possible \cite{BKP,AG}, and
the above results for the Hall resistivity can be tested.
Note that, provided the resonance (\ref{reso}) is close to the Fermi
energy $E_F$, an interesting effect may apear because
the Hall resistivity becomes very sensitive to the changes of $E_F$.

\section{Persistent current of free electrons in the plane
pierced by a flux tube}
\label{sec:pers}
\begin{figure}
\centerline{\epsfxsize=11cm \epsfbox{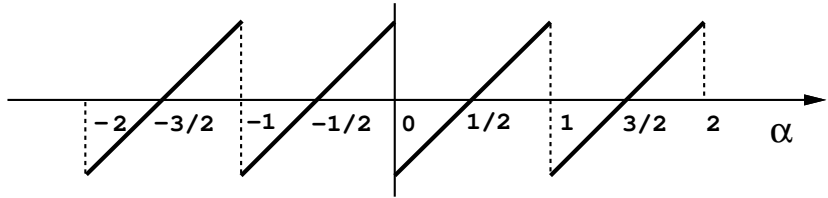}}
\caption{Characteristic dependence of a persistent current
 of free spinless fermions in the plane on the flux $\alpha$
(arbitrary units in the $y$ direction).}
\label{figper}
\end{figure}

\begin{figure}
\centerline{\epsfxsize=11cm \epsfbox{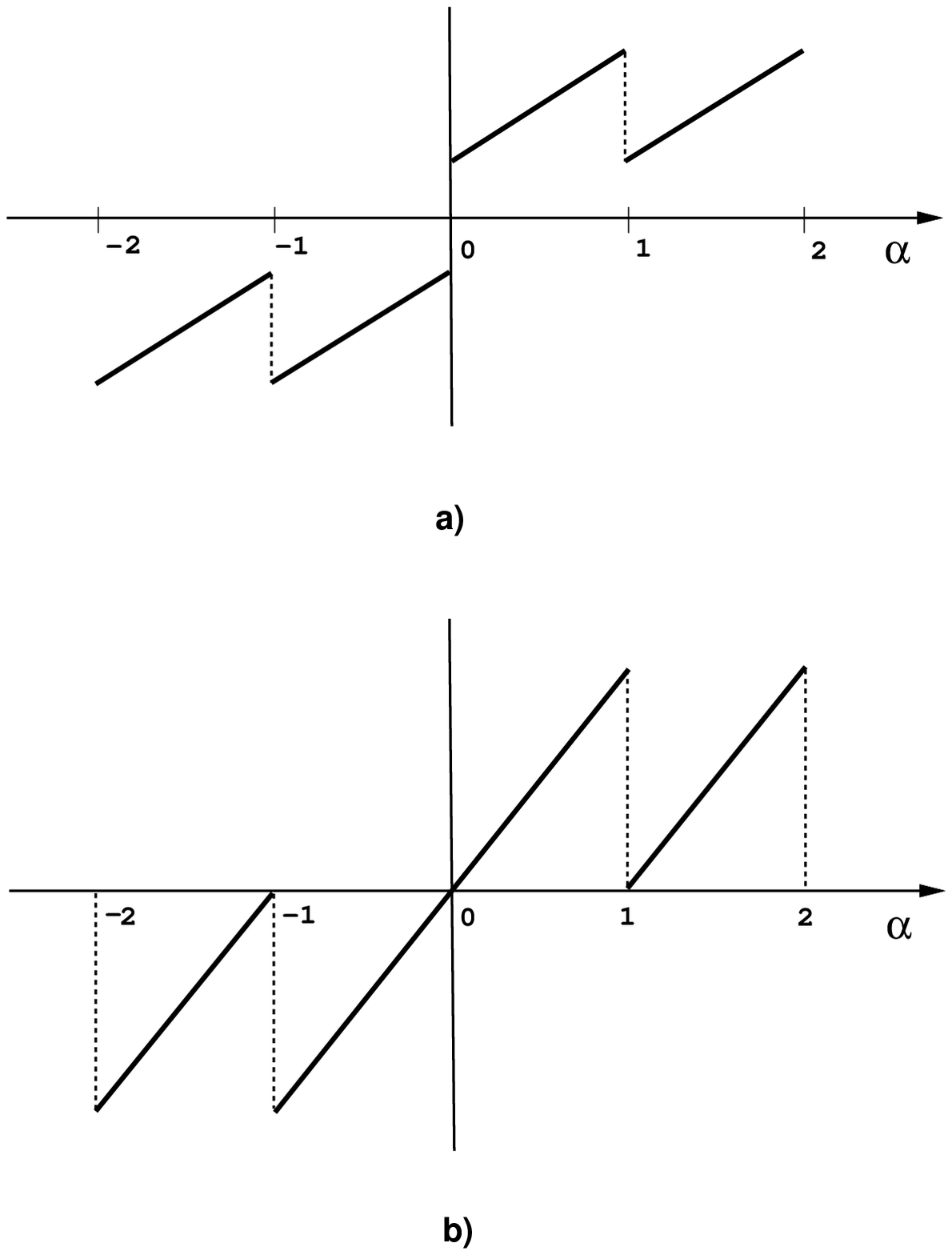}}
\caption{(a) The contribution of spin-up electrons to the persistent current
as a function of the flux $\alpha$, provided that the  phase-shift flip
occurs in a single channel.
(b) Persistent current of spin one-half fermions as a function of the flux $\alpha$.}
\label{figperf}
\end{figure}

The persistent current in a finite (ring) geometry 
was first
discussed in Ref.\ \cite{BIL}. It is reminiscent of the edge currents
(see discussion in Ref.\ \cite{BH} on their existence)
that arise in the presence of a magnetic field
in the absence of the electric force.
In the case of a bounded system, the energy levels are discrete
and the persistent current induced by a flux tube with flux $\Phi$,
carried by the $j$th eigenstate, is \cite{BIL}
\begin{equation}
i_j=-\frac{dE_j (\Phi)}{d\Phi}\cdot
\label{perse}
\end{equation}
In the case of an unbounded system the spectrum will 
have both a discrete and 
a continuous part. The persistent current carried
by an isolated eigenstate (from a point spectrum) is still 
given by formula
(\ref{perse}). The contribution of scattering states to a 
persistent 
current is then determined by the formula
\begin{equation}
dI(E,\alpha) =
(2\pi i)^{-1}\partial_\Phi
\left[\ln\det\mbox{S}(E,\Phi)\right]dE,
\label{aaa}
\end{equation}
derived by Akkermans {\em et al.}\ \cite{AA}.
Here  S$(E,\Phi)$ is the on-shell
scattering matrix in the presence of a flux tube,
and  $dI(E,\Phi)$ is the differential contribution to the
persistent current at energy $E$.
The persistent current was defined with respect to a point. It was
given by the total current through a line that extends from that point
to infinity, in the absence of currents through the external leads.
Now, by the Krein-Friedel formula (\ref{krein}),
$\ln\det\mbox{S}(E,\Phi)$ is directly related to the
change $\triangle N_\alpha(E)$ of the IDOS, and hence
(by using $\Phi=\alpha\Phi_0$) we have
\begin{equation}
dI(E,\alpha) = \frac{e}{hc}\,\partial_\alpha
\left[\triangle N_\alpha(E)\right] dE.
\label{aaa1}
\end{equation}
As has been shown in Sec.\ \ref{sec:krein},
$\triangle N_\alpha(E)$ is a symmetric function of $\alpha$.
Therefore, a persistent current is an antisymmetric function of
$\alpha$ (see Figs. \ref{figper} and \ref{figperf}).
One can show that the formula (\ref{aaa1}) reduces to
(\ref{perse}) in the case of the discrete spectrum, and  in fact, 
the formula is valid for both  continuous and discrete parts of the
spectrum. In the latter case, the DOS is formally given by
$\rho(E,\Phi)=\sum_j \delta [E-E_j(\Phi)]$, where the summation is over
all discrete levels. Hence
\begin{equation}
\frac{d\rho(E',\Phi)}{d\Phi}=-\sum_j
\delta'[E-E_j(\Phi)]\,\frac{dE_j}{d\Phi}
\cdot
\label{dddo}
\end{equation}
Therefore,
\begin{equation}
\int^E \frac{d\rho(E',\Phi)}{d\Phi}\,dE'=-\sum_{j,E_j<E}
\delta[E-E_j(\Phi)]\,\frac{dE_j}{d\Phi}\cdot
\label{ddd1}
\end{equation}
Now, by substituting the result into Eq.\ (\ref{aaa1})
and after integrating up to the Fermi energy $E_F$, one recovers
the sum over all contributions of single levels, 
as given by the formula (\ref{perse}), below the Fermi energy $E_F$.

For spinless fermions neither bound states nor a phase-shift flip occur
and the change in the DOS is given by (\ref{denreg}).
Therefore, when all states
below the Fermi energy $E_F$  are occupied,
one finds \cite{CMO1} that the persistent current of spinless fermions
around the origin, which is pierced by a flux tube, is
\begin{equation}
I=\frac{eE_F}{hc}\,(\eta-1/2).
\label{percles}
\end{equation}   
The current depends linearly on $\eta$ (cf. Ref.\ \cite{AA}
where it is a constant) as  it does in small one-dimensional metal 
rings \cite{CG}.  In contrast to Ref.\ \cite{AA},
the current is antisymmetric not only about the values
$\alpha=n$, where $n$ is an integer, but also
about values  $\alpha=n+1/2$ where it vanishes
(see Fig.\ \ref{figper}). 
The  latter values of $\alpha$ are such as the former the values
of $\alpha$ where time invariance is preserved.

In the case of spin one-half fermions, the contribution of spin-up
fermions to the persistent current is still given by formula
(\ref{percles}). The contribution of  spin-down fermions
depends on their magnetic moment $g_m$. 
At the critical value
of $g_m=2$, either the phase-shift flip or a bound state can occur.
The most general expression  for the contribution
of scattering states to $\triangle N_\alpha(E)$ for $E\geq 0$, which 
includes the situation  where  bound states or a phase-sfhift flip
are present, 
is given by  Eq. (\ref{intsing}). The persistent current in this case
is obtained by substituting the result (\ref{intsing}) for
$\triangle N_\alpha(E)$ directly in  Eq. (\ref{aaa}).
Here, one must not forget that the bound state energies
also depend on flux \cite{AG}. 
In the case that the phase-shift flip occurs in the $l=-n$ channel,
the contribution of spin-down fermions to the persistent current
is given by the formula 
\begin{equation}
I=\frac{eE_F}{hc}\,(\eta+1/2)
\label{percup}
\end{equation} 
[see Fig.\ \ref{figperf}(a)].
Therefore, the total persistent current of spin one-half fermions
in the plane when a phase-shift flip occurs is
\begin{equation}
I=2\frac{eE_F}{hc}\,\eta
\label{perctot}
\end{equation} 
[see Fig.\ \ref{figperf}(b)]. This is another important difference
from Ref.\ \cite{AA}, where they obtained a result that the 
total current is zero.
The result is consistent with general 
requirements \cite{AA} of periodicity with regard to 
$\alpha\rightarrow\alpha\pm 1$, and antisymmetry
with respect to $\alpha\rightarrow -\alpha$.
Provided that the phase-shift flip occurs both in the $l=-n$ and 
$l=-n-1$ channels, the contribution of spin-down fermions
to the persistent current is the same as that of spin-up
fermions. However, as has been extensively discussed 
in Secs. \ref{sec:reg}
and \ref{sec:ren}, the latter case of the phase-shift flip in two
channels is less probable than that of the phase-shift 
flip in a single channel.

Observations
of the persistent current may finally reveal the resonance
(\ref{reso}) in the AB scattering,
since near it the current becomes very sensitive to the change of
the Fermi energy $E_F$ and of the flux.
The similarity between the scattering in the presence of the 
AB potential and in the field of a cosmic string naturally suggests 
that a similar current should occur in the field of a cosmic string, 
too. 

\section{The 2nd virial coefficient of nonrelativistic any\-ons}
\label{sec:vir}
Anyons are usually represented as either bosons or fermions 
threaded by the flux tube with the flux $\Phi$. 
Noninteracting anyons are described by the Hamiltonian
\begin{equation}
H= \sum_{\stackrel{i,j}{\scriptscriptstyle i\neq j}}^N 
\frac{({\bf p}_i-\frac{e}{\hbar c}{\bf A}_{ij})^2}{2m},      
\end{equation}
where $N$ is the number of anyons, and 
${\bf A}_{ij}={\bf A}({\bf r}_i-{\bf r}_j)$ 
is nothing but the AB potential (\ref{abpot})
centered at the position of the $jth$ particle,
\begin{equation}
{\bf A}_{ij}=\frac{\Phi}{2\pi |{\bf r}_i -{\bf r}_j|^2}\, 
{\bf z}\times ({\bf r}_i -{\bf r}_j)=-{\bf A}_{ji},      
\end{equation}
written here in a slightly different form with ${\bf z}$ the
unit vector in the direction of the flux.
In what follows we shall consider anyons in the presence 
of a pairwise  interaction $U({\bf r} -{\bf r}')$,
\begin{equation}
U({\bf r} -{\bf r}')= -\lambda\frac{2\pi\hbar^2}{m}\alpha\,
\delta^{(2)}({\bf r} -{\bf r}')=
-\lambda\frac{\hbar^2}{m}\frac{\alpha}{|{\bf r} -{\bf r}'|}
\,\delta(|{\bf r} -{\bf r}'|).
\label{singi}
\end{equation} 
By transforming as usual 
to the center of mass and relative coordinates $(r,\varphi)$, and 
leaving aside the free motion of the center of mass, 
the relative Hamiltonian takes the form \cite{CGO,MT,LM}
\begin{equation}
H_{rel}=-\frac{\hbar^2}{m}\left[\partial_r^2+\frac{1}{r}\partial_r
-\frac{1}{r^2}(-i\partial_\varphi+
\alpha)^2\right]+U({\bf r} -{\bf r}'),
\end{equation}
where, as above, $\alpha=\Phi/\Phi_0$.
The form of the relative Hamiltonian $H_{rel}$ corresponds  \cite{YSW}
to that used in Refs. \cite{AD} and \cite{BHR}.
 One neglects the electrostatic
forces between anyons by assuming the limit $e\rightarrow 0$
with $\alpha$ fixed.
The relative wave function is parametrized as $\exp(iL\varphi)\,f(r)$,
where $L$ is standard, namely $L=2l$ for bosons and $2l+1$ for fermions.
After parametrization, the relative Hamiltonian $H_{rel}$ takes 
a form similar to that of $H_l$ (\ref{schrham}), 
provided
the substitutions $\lambda=g_m$ and $s_z=-1$ are made, and the
reduced mass $\bar{m}= m/2$ is used. 
Due to the parametrization of the relative wave function,
the parameter $\nu$ is now
\begin{equation}
\nu =2l+\alpha
\end{equation}
provided one starts from the bosonic end, or
\begin{equation}
\nu=2l+1+\alpha
\end{equation}
when one starts from the fermionic end \cite{CGO,LM}.

The equation of state of a real gas expanded in 
powers of the density, $\mu=N/V$, is
\begin{equation}
PV=\frac{N}{\beta}(1+a_2\mu +a_3\mu^2 +\ldots\,)
\end{equation}
where the $a_j$ stand for the virial coefficients (see, for example, 
Ref.\  \cite{OT}). Here, $P$ is the pressure, $V$ is the volume,  
and $\beta=1/kT$.
The calculation of $a_2(T)$ only requires a knowledge
of two-body interaction.
We shall show that the results of preceding sections
(with a slight modification)
can be directly applied to the calculation of the $2nd$
virial coefficient $a_2(T)$ of the gas of anyons. 
Let us first consider the noninteracting case $\lambda=0$.
By using the Krein-Friedel formula one can calculate the
change of the DOS in a way similar to that used in
 Sec. \ref{sec:krein}. 
In the case of bosons one finds
\begin{eqnarray}
\lefteqn{\ln\det\mbox{S}=\sum_{l=-\infty}^\infty 2i\delta_l
= i\pi \sum_{l=-\infty}^\infty (|2l|-|2l+\alpha|)
}\nonumber\\
&&
 =2\pi i 
\left.\left[ \sum_{l=1}^\infty l^{-s}
-\sum_{l=0}^\infty (l+\eta_a)^{-s}
-\sum_{l=1}^\infty (l-\eta_a)^{-s}\right] \right|_{s=-1}
\nonumber\\
&&
=2\pi i  \left.\left[
2\zeta_R(s)-\zeta_H(s,\eta_a)-\zeta_H(s,1-\eta_a)\right]
\right|_{s=-1}\nonumber\\
&& = -2\pi i\, \eta_a(1-\eta_a),
\end{eqnarray}
and hence
\begin{equation}
\triangle\rho_\alpha^b(E)=-\eta_a (1-\eta_a)\delta(E).
\label{densb}
\end{equation}
Now, in the case of anyons, $\eta_a$ is the fractional part of $\alpha/2$.
Note that formally, if the sum in Eq. (\ref{frsum})
is restricted to even $l$ only, the result (\ref{densb}) is 
twice that in 
the AB potential with unrestricted $l$.  
In the case of fermions (here $\eta_a$ is 
supposed to be within the range $|\eta_a|<1/2$),
\begin{eqnarray}
\lefteqn{\ln\det\mbox{S}=\sum_{l=-\infty}^\infty 2i\delta_l
= \pi i\sum_{l=-\infty}^\infty (|2l+1|-|2l+1+\alpha|)=
}\nonumber\\
&&
 =2\pi i
\left.\left[ \sum_{l=0}^\infty (l+1/2)^{-s}
+\sum_{l=1}^\infty (l-1/2)^{-s}
-\sum_{l=0}^\infty (l+\eta_a+1/2)^{-s}
-\sum_{l=1}^\infty (l-\eta_a -1/2)^{-s}\right] \right|_{s=-1}
\nonumber\\
&& =
2\pi i \left.\left[
2\zeta_H(s,1/2)-\zeta_H(s,\eta_a+1/2)-\zeta_H(s,1/2-\eta_a)\right]
\right|_{s=-1} = 2\pi i\, \eta_a^2,
\end{eqnarray}
and hence
\begin{equation}
\triangle\rho_\alpha^f(E)=\eta_a^2\, \delta (E).
\label{densf}
\end{equation}
Now, the two-body interaction partition function
$Z_{int}(\beta)$ can be calculated from
\begin{equation}
Z_{int}^{b,f}(\beta) =\int e^{-\beta E}\, 
\triangle\rho_\alpha^{b,f}(E)\, dE.
\label{defz}
\end{equation}                               
Note that $Z_{int}(\beta)$ vanishes for $\eta_a=0$ when 
interactions (including the AB interaction) are switched off.
The integration here runs over the whole spectrum. However, since
there are no bound states, 
$\triangle\rho_\alpha(E)$ is zero for $E<0$ and 
the integral reduces to the Laplace transform.
By inserting Eqs. (\ref{densb}) and (\ref{densf})
into  Eq. (\ref{defz}) one finds that the partition functions
do not depend on the temperature $T$ \cite{CGO},
\begin{equation}
Z_{int}^b=-\eta_a(1-\eta_a), \hspace*{1cm}Z_{int}^f=\eta_a^2.
\label{zres}
\end{equation}
$a_2(T)$  can be directly expressed in terms of 
the two-body partition function and has the form
\begin{equation}
a_2^b(T)=-\frac{\pi\beta}{2m}
\left[1+8 Z_{int}^b\right]
\label{2vbdef}
\end{equation}
in the case of bosons, and
\begin{equation}
a_2^f(T)=\frac{\pi\beta}{2m}\left[1-8 Z_{int}^f\right]
\label{2vfdef}
\end{equation}
in the case of fermions \cite{CGO}. The prefactor in the last 
two expressions can be written as $\lambda_T^2/4$, where
$\lambda_T= \sqrt{2\pi\beta/m}$ is the thermal length.
Now Eqs. (\ref{zres}), (\ref{2vbdef}), and (\ref{2vfdef}) imply
that the $2$nd virial coefficients are periodic with respect
to $\eta_a\rightarrow\eta_a\pm1$, i. e., with respect to
$\alpha\rightarrow\alpha\pm2$ \cite{AD}. 
In case  of bosons one finds for $0<\alpha<2$ that
\begin{equation}
a_2^b(T)= -\frac{\lambda_T^2}{4}
\left(1-4\alpha+2\alpha^2\right)=
\frac{\lambda_T^2}{4}
\left[1-2(\alpha-1)^2\right].
\label{bvir}
\end{equation}
Similarly, in the case of fermions one obtains
for $-1<\alpha<1$ that
\begin{equation}
a_2^f(T) =\frac{\lambda_T^2}{4}\left(1-2\alpha^2\right)=
-\frac{\lambda_T^2}{4}\left[1-4(\alpha+1)+2(\alpha+1)^2\right].
\label{fvir}
\end{equation}
The result for other ranges of $\alpha$ is obtained by using 
the periodicity.
The $2$nd virial coefficients $a_2^b(T)$ and  $a_2^f(T)$ are written 
in two equivalent forms. The second form clearly shows that 
if $\alpha$ is raised to $\alpha+1$, then $a_2^b(T)\rightarrow a_2^f(T)$.
Similarly, if $\alpha$ is lowered by $1$, then 
$a_2^f(T)\rightarrow a_2^b(T)$.

So far, we have only reproduced the known results
for the $2nd$ virial coefficients of noninteracting 
anyons \cite{CGO,AD}. In the presence of an interaction
(as  may be the case for anyon-antianyon interaction)
the virial coefficients will change. The results for the
DOS in the AB potential then enable us to calculate
them in the special case of the potential $U({\bf r}-{\bf r}')$.
The analysis is similar to that of Sec. \ref{sec:krein}.
Therefore, provided the parameter $\lambda\neq 2$ in  Eq. (\ref{singi}), 
no bound state is present in the spectrum and the $2nd$
virial coefficients $a^{b,f}_2(T)$ are still given, respectively,
by Eq. (\ref{bvir}) or Eq. (\ref{fvir}).
They exhibit quite a lot of rigidity with respect to the interaction
and they start to change only at the critical coupling 
$\lambda_c=2$ when the phase-shift flip or a
bound state  may occur (at an energy that depends on the details 
of the limit when the radius of the flux tube shrinks to zero).
The only change with regard to the AB scattering discussed in
previous sections is that the parameter $\nu$ changes
here by $2$ and hence a bound state or the phase-shift flip
can only occur in a {\em single} channel. 
Then the change of the IDOS $\triangle N_\alpha(E)$ is 
given either by
\begin{equation}
\triangle N_\alpha(E) = {\cal N}(\eta_a)
+\frac{1}{\pi}\arctan
\left(\frac{\sin(\eta\pi)}{\cos(\eta\pi)-(|E_{-n}|/E)^{\eta}}
\right)
\end{equation}
or
\begin{equation}
\triangle N_\alpha(E) =  {\cal N}(\eta_a) 
-\frac{1}{\pi}\arctan
\left(\frac{\sin(\eta\pi)}{\cos(\eta\pi)+
(|E_{-n-1}|/E)^{(1-\eta)}}\right),
\label{intansing}
\end{equation}
where either ${\cal N}(\eta_a)=-\eta_a(1-\eta_a)$ or 
 ${\cal N}(\eta_a)=\eta_a^2$
depending on $n$ and whether one starts from the bosonic or the fermionic 
end. To calculate $Z_{int}(\beta)$ in this case one integrates
by parts the general formula (\ref{defz}) and 
rewrites it in terms of the change $\triangle N_\alpha(E)$ of the 
IDOS \cite{Ex}. Now a bound state is present and one obtains 
in general
\begin{equation}
Z_{int}(\beta)= \sum_{b}e^{-\beta E_b}
+
\beta \int_0^\infty e^{-\beta E} \triangle N_\alpha (E)\, dE,
\label{genz}
\end{equation}                               
where the sum here in principle runs over all bound states. 
Note that at the critical coupling,
the partition function depends on $T$.
Eventually, the $2nd$ virial coefficients are obtained
by inserting the result in  either (\ref{bvir}) or (\ref{fvir}).
In Table I the results are presented
for the partition function, and the $2$nd virial
coefficient for bosons and fermions in the case when 
the phase-shift flip occurs for  $n=[\alpha]$ irrespective of whether
odd ($n=n_o$) and  even ($n=n_e$). 
The results are presented in terms of $\eta$.
Provided $n$ is even, $\eta$ is 
added to (subtracted from) $Z_{int}(\beta)$ in the case of bosons
(fermions). The reason is that in the former case the condition
$0<|\nu|<1$ is satisfied for the only $l$  given by $2l=-n$ 
($2l=-n-2$), which implies that the parameter 
$\nu$ takes the value $\eta$ [$-(1-\eta)$].
When $n$ is odd, the situation is reversed and $\eta$ is 
subtracted from (added to) $Z_{int}(\beta)$  in the case of bosons 
(fermions). The calculation can be performed straightforwardly,
but care has to be taken with regard to the range of $\eta_a$:
$0<\eta_a<1$ in the case of bosons and $|\eta_a|<1/2$ 
in the case of fermions. 
\begin{center}
TABLE I. The partition function $Z_{int}(\beta)$
and the $2$nd virial coefficient for bosons and fermions 
in the presence of the phase-shift flip for 
$n=[\alpha]$ odd and even.\vspace*{0.3cm} \\
\begin{tabular}{||c|c|c|c|c||} \hline\hline
 & & & &  \\
n &  $Z^b_{int}(\beta)$  & $a_2^b(T)$ &
 $Z^f_{int}(\beta)$ 
& $a_2^f(T)$ \\
& & & &   \\  \hline\hline
& &  & & \\ 
even & $\frac{1}{4}(2\eta+\eta^2)$  & $\frac{\lambda_T^2}{4}
\left[1-2(\eta+1)^2\right]$ & $\frac{1}{4}(-4\eta+\eta^2)$ & 
$\frac{\lambda_T^2}{4}
\left(1+8\eta-2\eta^2\right)$
\\
& & & &  \\  \hline
& &  & & \\ 
odd &  $-\frac{1}{4}(1+4\eta-\eta^2)$ & $\frac{\lambda_T^2}{4}
\left(1+8\eta-2\eta^2\right)$&  $\frac{1}{4}(\eta+1)^2$  &
$ \frac{\lambda_T^2}{4}
\left[1-2(\eta+1)^2\right]$ \\
& & & &  \\ \hline\hline
\end{tabular}
\end{center}
The results presented in Table I, apart from the periodicity 
of the $2$nd virial coefficients, demonstrate clearly that the $2$nd virial coefficients interpolate between the 
fermionic and bosonic case as $\alpha$
changes by unity. Namely, the $2$nd virial coefficient in the $n_oB$ case
(bosonic end with $n=n_o$ odd) is the same as that in
the $n_eF$ case (fermionic end with $n=n_e$ even), and similarly
the result  for the $2$nd virial coefficient in the $n_oF$ case
is the same as that in the $n_eB$ case. The reason is that
in the respective cases, the partition functions $Z^{b,f}_{int}$
receive the same contribution. This is not restricted to the case
when the contribution is given by the phase-shift flip only, but
holds in the general case [(\ref{intansing}) and (\ref{genz})]
when the bound state is present.
Then, according to Eqs.\ (\ref{2vbdef})
and (\ref{2vfdef}), the changes in the $2$nd virial coefficients
are identical. Therefore, the interpolation
between different statistics as $\alpha\rightarrow\alpha\pm1$
holds both with and without the short-range interaction
(\ref{singi}).
There is, however, one distinguishing feature of the $2$nd virial 
coefficients at the critical coupling observed in Ref.\ \cite{BHR},
namely, their {\em discontinuity}. Since a discontinuity in the
virial coefficient implies a discontinuity in the free energy,
to the same order of approximation this points to some kind of
phase transition \cite{BHR}. It has  already been mentioned that at the
critical coupling $\lambda_c=2$, the scale invariance is generally
broken and one expects a different physics in this case.

Our results cover naturally the case of the spin one-half 
anyons \cite{BHR}, too. In the latter if
we ignore the anomalous part of the magnetic moment, but add
a Zeeman interaction, an anyon-anyon potential 
$U({\bf r}-{\bf r}')$ arises which has  the form
\begin{equation}
U({\bf r}-{\bf r}')= \frac{\hbar^2}{m}(s_1+s_2)
\frac{\alpha}{|{\bf r}-{\bf r}'|}
\,\delta(|{\bf r}-{\bf r}'|),
\label{urint}
\end{equation}
where $s_1=\pm 1$ and $s_2=\pm 1$ are spin projections 
on the direction of the flux tube \cite{BHR}.
Provided that at least one of the spins does not have an orientation
opposite to the orientation of the magnetic field, then, 
depending on the statistics, the result for the  $2$nd virial 
coefficient is given either by Eq.\ (\ref{bvir}) or  Eq.\ (\ref{fvir}),
in accord with \cite{BHR} and with previous results \cite{CGO,AD}:
the interaction does not have its critical value and one should
recover the noninteracting case.
If both spins are opposite to the direction of the magnetic field,
the critical potential with $\lambda_c=2$ arises and, provided the
phase-shift flip takes the place, the results for the $2$nd virial 
coefficient of anyons can be read off from Table I. 
Our result in the $n_oF$ case agrees with that of 
Blum {\em et al.} \cite{BHR}. In the $n_eF$ case they did not take into
account that the phase-shift flip may occur as they assumed
that in the AB scattering the phase-shift flip cannot occur in the
$l\!=\!-n-1$ channel. Therefore, not
surprisingly, in the absence of the phase-shift flip 
in Ref.\ \cite{BHR} they obtained the same result
(\ref{fvir}) for the $2$nd virial coefficient
in the $n_eF$ case as in the absence
of the interaction $U({\bf r}-{\bf r}')$. 
As has been discussed in Sec.\ \ref{sec:reg}, it
 is more difficult to create a bound state in the $l=-n-1$
channel, and the phase-shift flip in this channel may possibly not
occur. If this happens, then our result for the $2$nd virial
coefficient coincides with that of Ref.\ \cite{BHR}.

For the case of an unpolarized system, the complete $\bar{a}_2^f(T)$
is obtained by averaging over four spin states.
Provided that there is no phase-shift flip in the $n_eF$ case,
 $a_2^f(T)$ is the same in all four cases and is given
by Eq.\ (\ref{fvir}). In the presence of the phase-shift flip, 
 one of the values
$a_2^f(T)$ is then given by Table I. In the $n_oF$ case, three values
of $a_2^f(T)$ are the same and given by Eq.\ (\ref{bvir}), because
of the interpolation between fermionic and bosonic statistics.
The remaining value of $a_2^f(T)$ under the presence of the phase-shift flip
is again taken from Table I.
Finally, one obtains
\begin{equation}
 \bar{a}^f_2(T)=\frac{\lambda_T^2}{4} \left\{
\begin{array}{ll}
(1-2\eta^2),\ &\ n_eF\ \mbox{case without the phase-shift flip}\\
(1+2\eta -2\eta^2),\ &\  n_eF\ \mbox{case with the phase-shift flip}\\
(-1+2\eta-2\eta^2),\ &\ n_oF\ \mbox{case}.
\end{array}
\right.
\end{equation}

It is worth noting that anyons with a short-range
pairwise {\em attractive} interaction 
occur in the nonrelativistic limit, taken up to the terms of order
$v^2/c^2$, of the topologically massive planar electrodynamics 
\cite{DJT} defined by the action
\begin{eqnarray}
\lefteqn{S=S_{matter}+ }\nonumber\\
&&
\frac{1}{L^2}\int \left(
-\frac{1}{4}F_{\alpha\beta}
F^{\alpha\beta}+\frac{1}{2R}\epsilon_{\alpha\beta\gamma}
A^\alpha\partial^\beta A^\gamma \right)dt\, d^2{\bf r},    
\label{planar}
\end{eqnarray}
with a nonminimal coupling \cite{St,Kog}
\begin{eqnarray}
\lefteqn{S_{matter}=S^0_{matter}-}\nonumber\\
&&
\sum_{a=1}^N
\int  J^\mu_{(a)}(x)
\left[\xi_a A_\mu(x)-\frac{1}{2}g_aL^{-2}
\epsilon_{\mu\nu\sigma}F^{\nu\sigma}\right]dt\,d^2{\bf r}.
\end{eqnarray}
Here, $|R|$ represents the screening length of the electromagnetic 
interaction, $L$ is the arbitrary scale parameter, $\xi_a$
stands for the $a$th particle's charge in units of $L$,
$J^\mu_{(a)}$ denotes the standard one-particle current normalized
to the unit charge, and $N$ is the number of the pointlike particles.
In the model the photon is massive with a mass $M=\hbar/c|R|$.
Note that $F^{12}$ is actually a magnetic field, 
and the nonminimal 
coupling term that is peculiar to $2+1$ dimensions 
\cite{St,Kog} is in this instance the familiar Pauli 
magnetic moment coupling that exists here 
even in the absence of a spin.
It has been shown by Stern \cite{St} that, provided
\begin{equation}
g_a=L^2 R\, \xi_a,
\label{gcrit}
\end{equation}
$\oint_C dx^i A_i$ becomes a topological invariant,
\begin{equation}
\oint_C dx^i A_i=gn.
\end{equation}
The screening length remains nonzero, and the electric charge and
magnetic moment balance each other in such a way that the radiation is 
{\em absent}. 
At the tree level, the theory at the critical coupling (\ref{gcrit})
can be exactly reexpressed as a simple {\em effective action 
at a distance} without the usual complications associated with 
the retardation, etc. \cite{St}.
The situation is analogous to that in pure
Chern-Simons (CS) theory,
\begin{equation}
S_{CS}=\frac{1}{2\Theta}\int \epsilon_{\alpha\beta\gamma}
A^\alpha\partial^\beta A^\gamma dt\,d^2{\bf r}    
\end{equation}
which is recovered in the limit $L^2\rightarrow\infty$, 
$R\rightarrow 0$, keeping fixed the parameter
\begin{equation}
\Theta=L^2 R.
\end{equation}                   
In the latter case the absence of radiation is 
simply because the photon gets infinitely heavy.
At the tree level, the theory at the critical coupling (\ref{gcrit})
can be exactly reexpressed as a simple {\em effective action 
at a distance} without the usual complications associated with 
the retardation, etc. \cite{St}.

\section{Discussion of the results and open questions}
\label{sec:conc}
The nonrelativistic scattering in the AB potential has been
analyzed. The DOS and scattering cross sections have been calculated
 and various applications have been discussed.
Despite the fact that the single-particle Green function is not a 
periodic function of $\alpha$,  it gives rise to 
the DOS which is a periodic and symmetric function of the flux
and it depends  only on the distance from the nearest integer.
It has been shown that the Krein-Friedel formula \cite{F}
is not restricted  to potentials of a finite range and 
can have a wider range of applicability. The Krein-Friedel formula may
be used even for long-ranged potentials when 
the sum over phase shifts is properly regularized.
In the case of the Aharonov-Bohm potential it is the
$\zeta$-function regularization that gives the correct answer.
By means of the Krein-Friedel formula the change of the
 DOS induced by the Aharonov-Bohm potential has been calculated
for different self-adjoint extensions which correspond to
different physics inside the flux tube. 
For the conventional setup, i.\ e., 
with zero boundary conditions at the
boundary of the flux tube, our result for the DOS (\ref{denreg})
confirms the anticipation of Comtet, Georgelin, and Ouvry
 \cite{CGO} that the change of the 
DOS is concentrated at the zero energy.
Whenever a bound state is present it is always accompanied by a
resonance at an energy proportional to the 
absolute value of the energy of the bound state.
In the presence of a bound state or a phase-shift flip, the symmetry
of the spectrum with respect to the change
$\alpha\rightarrow\alpha\pm 1$ is broken.
Nevertheless, scattering cross sections continue to be periodic
with respect to the substitution $\alpha\rightarrow\alpha\pm 1$.
                                                                
In the case of a regular flux tube of a finite radius $R$,
the question of the number of bound states has been clarified.
It has been shown that the number depends  not only on the
total flux but also on the energy of magnetic field.
It means that one has to be careful
in the choice of the regularization 
when discussing the physics with anomalous magnetic 
moment $g_m\geq 2$ or with a short-range interaction inside
the flux tube. When the AB vortex arises by some physical
mechanism such as vortices in the superconductor of type II do, then
the profile of a magnetic field is determined by the condition of
the minimum of energy. 
Our considerations have only been quantitative. The open question
which remains is what is the qualitative dependence on the energy
and whether (and if then, how) the number of bound states 
depends on higher moments $\int B^n(r)\,d^2\,{\bf r}$
of the magnetic field $B(r)$, where $n\geq 3$.
We have shown that a bound state in the $l\!=\!-n-1$
channel can exist provided $1-\eta$ is sufficiently small.
In the discussion of the $R\rightarrow 0$ limit, the existence of
a critical coupling $g_m=2$ was established. Provided the coupling
with the interior of the flux tube is smaller or greater than
the critical one, or when the coupling is not renormalized,
the limit coincides with that of the impenetrable flux tube.
Thereby, the result of Ref.\ \cite{AU} for $g_m=0$
is generalized.
The Aharonov-Casher theorem about the number of zero modes
has been corrected. It has been shown that they merge with
the continuous spectrum in the $R\rightarrow 0$ limit.
The origin of the phase-shift flip has been investigated.
It was  shown that the phase-shift flip \cite{Hag} may 
occur even in the absence of spin. 
The conditions for the phase-shift flip formulated in
Refs. \cite{Hag,MRW} have been found necessary but not sufficient.
In particular, provided  no bound state is in the $l\!=\!-n-1$ channel
for $R\neq 0$, then the phase-shift flip will not occur at the
$R\rightarrow 0$ limit.

A quantum-mechanical and nonrelativistic criterion of stability 
(\ref{instab}) in $2+1$ dimensions has been  discussed 
that only involves fundamental parameters of matter.
Despite that this criterion was derived under very 
restrictive hypotheses, it was shown to be compatible
with other, quantum-field theory treatments \cite{YH,VG}.
An open question remains regarding what will happen if the dynamics
of the shrinking of the radius of a flux tube is included.
Another open question is what field configuration actually
optimalizes the right-hand side of Eq.\ (\ref{instab}), i.\ e.,
what is the infimum of the right-hand side of Eq.\ (\ref{instab}) with
respect to variations of magnetic field $B(r)$, subject
to the constraint (\ref{bconst}).

It has been shown that in the presence of bound states or
a phase-shift flip, the differential scattering cross 
section becomes asymmetric
with respect to the substitution $\varphi\rightarrow -\varphi$,
and gives rise to the Hall effect. The Hall resistivity has
been calculated [see Eq.\ (\ref{reshall})]. 
The persistent current of free electrons in the plane has been
discussed and the results of Ref.\ \cite{AA} have been extended.
In contrast to  Ref.\ \cite{AA}, we have found that the total
persistent current in the spin one-half case is different from zero,
and the persistent current of  both the spinless and 
the spin one-half fermions
depends linearly on $\eta$.
The formal similarity between
the scattering of electrons in the AB potential and that in the
field of a cosmic string suggests that a persistent current
will also appear in the latter case. 

The known results for the 
$2nd$ virial coefficients  of the nonrelativistic anyons \cite{CGO,AD}
were reproduced and their values  were
calculated for the case when anyons interact
via a short-range attractive potential,
$U({\bf r}-{\bf r}')$, proportional to the Dirac $\delta$ function 
given by  Eq.\ (\ref{singi}). 
The case considered here is complementary to the case
of a long-range potential $g/r^2$, considered
by Loss and Fu \cite{LF}. In the present case,
the $2nd$ virial coefficients  were shown to be remarkably 
stable when such an interaction is switched-on.
They do not change when the interaction is switched on
until the coupling constant reaches its critical strength.
At the critical coupling
the $2$nd virial coefficients become {\em discontinuous}
as a function of $\alpha$ but their periodicity
with respect to $\alpha\rightarrow\alpha\pm 2$ and
the interpolation $a_2^{b,f}(\alpha\pm 1)=a_2^{f,b}(\alpha)$
between different statistics still hold. 
These results were obtained by using the $\zeta$-function regularization 
which has an obvious advantage with regard to other finite box
or harmonic potential regularizations.
By using the $\zeta$-function regularization one avoids the necessity of a discretization of energy levels
for computing the partition function.
Our results
naturally generalize those of Blum {\em et al.} \cite{BHR}
for nonrelativistic spin one-half anyons for the case of an anomalous
magnetic moment and when a bound state occurs in the spectrum.

Our results have been formulated in terms of self-adjoint extensions.
A self-adjoint extension is the rigorous limit $R\rightarrow 0$.
Therefore, in the case of a flux tube of finite radius $R$,
our results can be applied when the radius $R$ is negligibly small
when compared to all other length scales in the system
under consideration. The parameters $\triangle_{-n}$ and 
$\triangle_{-n-1}$ of the relevant  self-adjoint extension are then
determined from bound state energies in the $l=-n$ and $l=-n-1$ channels.
Hence, to check some of our results experimentally it is not necessary
to use a singular magnetic field. By using the duality
discussed in Sec. \ref{sec:ren} the physics in the presence
of an almost singular flux tube can be tested in experiments
with slow electrons scattered off a general rotationally invariant
two-dimensional magnetic field
$B(x,y)$ that obeys the finite-flux condition (\ref{bconst}).
From the experimental point of view we stress the
following experiments that should be performed:

\begin{itemize}
\item Measurement of the phase shift flip in the $l\!=\!-n-1$ channel.
Depending on the flux it may or may not occur in this channel.

\item  Measurements of the Hall resistivity either in 
the dilute vortex limit or for the single vortex.

\item  Measurements of the resonance.
\end{itemize}
A hypothetical setup to observe the resonance is
to study the transmission through a single flux tube.
A realistic physical realization is that suggested originally by
Rammer and Shelankov \cite{RS} and later realized experimentally
by Bending, Klitzing, and Ploog \cite{BKP} (see Fig.\ \ref{prfgrs}).
When homogeneous magnetic field is switched on, the conventional
superconductor is penetrated by vortices of flux with $\alpha=1/2$,
i.\ e., exactly that flux at which the resonance is at infinity.
Recent measurements on YBCO-delta rings with three grain-boundary
Josephson junction \cite{Kir}, however, reported the observation
of vortices  that carry a flux $\alpha=1/4$ which is {\em smaller} 
than the standard flux quantum $h c/2e$ 
(corresponding to $\alpha=1/2$) in the superconductor.
Therefore, when the  high-T${}_c$ YBCO film is used
as a gate  on top of the heterostructure containing 2DEG,
the resonance is at some finite energy and 
could in principle be observed.
Another possibility is to measure energy dependence 
of phase shifts and
(as a precursor of the resonance) the time delay (\ref{tdelay}).

An interesting open problem concerns the shape of the resonance 
(\ref{resshape})
in the AB scattering that is not of the Breit-Wigner form.
The latter is a general consequence of the {\em analyticity} of
scattering amplitudes that is usually taken as an equivalent to
{\em causality}. However, the rigorous proof of the
equivalence requires precise localization in time and  
in the energy for the incident wave packet.
This is, however, impossible due to the uncertainity relations
impossible \cite{EOP}. We have undertaken the
 analysis of the problem 
in terms of the Jost functions. However, we have not found
the origin of this behavior and we postpone
the solution to this problem. 
The discussion  might in principle
shed some light on the analyticity principle usually adopted
in the axiomatic quantum-field theory as a substitute for
causality.

Another interesting problem is to find the analog of the Levinson
theorem (see Ref. \cite{RN}, p. 356, and Ref. \cite{ReS}) and the 
generalization of Bargmann's
inequalities \cite{Se} for singular potentials. In the case of
regular potentials of finite range one knows, thanks to this
theorem, that if the phase shifts 
are normalized so that $\delta_l(E)=0$ for $E\rightarrow \infty$, then
$\delta_l(0)$ gives the number of bound states in the channel $l$.
In the case of the AB potential 
the phase shifts (\ref{convshift}) essentially do not depend
on the angular momentum $l$ and on the energy $E$.
 In the case of $\alpha<1$ they are 
either $-\pi\alpha$ for positive $l$ or $\pi\alpha$ 
for negative $l$. It is impossible in general
to normalize phase shifts in the singular case in the same way as for 
regular potentials, since they do not depend on the energy.

\section{Acknowledgments}
I should like to thank IPN at Orsay for warm
hospitality during my stay there and, in particular, 
A. Comtet, Y. Georgelin, M. Knecht, S. Ouvry, and J. Stern
for suggestions on literature and many useful 
and stimulating discussions, and Mme. C. Paulin
for drawing the figures. Partial support by  the UK EPSRC Grant 
GR/J35214 and by the Grant Agency 
of the Czech Republic under Project No. 202/93/0689  are 
also gratefully acknowledged.
Last but not least I thank R. C. Jones
for careful reading of the manuscript.    
\appendix

\section{Some useful integrals}
\label{app:int}
To perform the integral in Eq. (\ref{int}) one uses the result
that
\begin{equation}
J_\nu(kr_y)=\frac{1}{2}\left[
H^{(1)}_\nu(kr_y)+H^{(2)}_\nu(kr_y)\right]
\end{equation}
and deforms the integration contour as in Fig. \ref{fig1}.
$H^{(1)}$ and $H^{(2)}(z)$ are the Hankel functions,
\begin{equation}
H^{(1)}(z)=\frac{i}{\sin \nu\pi}\left[e^{-\nu\pi i}J_\nu(z)
-J_{-\nu}(z)\right].
\end{equation}
For real arguments,
\begin{equation}
H^{(1)}(x)=H^{(2)}(x)^*
\end{equation}
[see Ref.\  \cite{AS}, relations (9.1.3) and (9.1.4)].                       
The integral with $H^{(1)}_\nu(kr_y)$ is now deformed
to the contour $C_1$ while that with $H^{(2)}_\nu(kr_y)$ 
is deformed to the contour $C_2$. 
Note that $H^{(1)}_\nu(z)$ 
[$H^{(2)}_\nu(z)$] is exponentially decreasing in the upper
(lower) half of the complex plane (see Appendix \ref{asymp}).
Since it is assumed that $r_x<r_y$ the integrand in Eq.  
(\ref{int}) is exponentially decreasing as well.
Now, the  contributions of any of the two integrals from the 
imaginary axis do not vanish. However,
by using the identities (9.1.35) and (9.1.39) from Ref.\  \cite{AS},
\begin{equation}
J_\nu(z e^{\pm\pi i})= e^{\pm\nu\pi i} J_\nu(z),
\end{equation}
\begin{equation}
H^{(2)}_\nu (z e^{-\pi i})=-e^{\nu\pi i} H^{(1)}_\nu(z),
\label{h2j}
\end{equation}
\begin{equation}
H^{(1)}_\nu (z e^{\pi i})=-e^{-\nu\pi i} H^{(2)}_\nu(z),
\end{equation}
one can show that
\begin{equation}
\int_{i\infty}^0 \frac{z dz}{q^2-z^2} J_\nu(zr_x) H^{(1)}_\nu(zr_y)=
-\int_{-i\infty}^0\frac{z dz}{q^2-z^2} J_\nu(zr_x) H^{(2)}_\nu(zr_y),
\end{equation}
i.\ e., the integrals on the imaginary axis cancel in the sum.
Thanks to (\ref{h2j}) one has for $\nu=0$
\begin{equation}
H^{(2)}_0 (z e^{-\pi i})=- H^{(1)}_0 (z),
\end{equation}
and the same applies to the integral in Eq. (\ref{gr:res}), too.
Once the integrals in Eqs. (\ref{gr:res}) and (\ref{int}) are 
represented as 
the sum of two contour integrals in the complex plane they can be 
simply taken by the residue theorem.

In order to calculate 
$\mbox{Tr}\,\triangle G_\alpha({\bf x},{\bf x},M)$
note that
\begin{eqnarray}
\lefteqn{
\mbox{Tr}\,\triangle G_\alpha({\bf x},{\bf x},M)= \int_{R^2}
\triangle G_\alpha({\bf x},{\bf x},M)\,d^2{\bf x} =}
\nonumber\\
&&
-\frac{\sin\eta\pi}{2\pi M^2}
\int_{-\infty}^\infty d\vartheta\,\int_{-\infty}^\infty d\omega
\frac{1}{(\cosh \omega +\cosh\vartheta)^2}
\,\frac{e^{\eta(\vartheta-\omega)}}
{1+e^{\vartheta-\omega}},
\end{eqnarray}
To perform the integral here
one makes the substitution \cite{Com}
\begin{equation}
x=\frac{\vartheta+\omega}{2},\hspace*{1cm} 
y=\frac{\vartheta-\omega}{2},
\end{equation}
with a Jacobian equal to $2$.
After some manipulations one finds
\begin{eqnarray}
\lefteqn{\int_{-\infty}^\infty d\vartheta\,\int_{-\infty}^\infty d\omega
\frac{1}{(\cosh \omega +\cosh\vartheta)^2}
\,\frac{e^{\eta(\vartheta-\omega)}}
{1+e^{\vartheta-\omega}}
=}\nonumber\\
&&\int_0^\infty\frac{dx}{\cosh^2 x}\int_{0}^\infty
\frac{\cosh(2\eta -1)y}{\cosh^3y} dy.
\end{eqnarray}
According to formula (3.512.1) of Ref.\ \cite{GR}
\begin{equation}
\int_0^\infty
\frac{\cosh(2\eta -1)y}{\cosh^3y} dy= 2 B(1+\eta, 2-\eta),
\end{equation}
where $B(x,y)$ is the Euler beta function,
\begin{equation}
B(x,y):=\frac{\Gamma(x)\Gamma(y)}{\Gamma(x+y)},
\end{equation}
and $\Gamma(x)$ is the Euler gamma function \cite{AS,GR}.
Since $B(1,1)=1$,
\begin{equation}
\int_0^\infty
\frac{1}{\cosh^2 x}\, dx= 1.
\end{equation}
Now, $\Gamma(3)=2$ and 
\begin{eqnarray}
\lefteqn{
2 B(1+\eta, 2-\eta)=\Gamma(1+\eta)\Gamma(2-\eta)=}\hspace*{3cm}
\nonumber\\
&&
\eta(1-\eta)\Gamma(\eta) \Gamma(1-\eta).
\end{eqnarray}
Therefore, by using (\ref{gamma1}),
\begin{equation}
\int_0^\infty
\frac{\cosh(2\eta -1)y}{\cosh^3y}\, dy= \pi\frac{\eta(1-\eta)}
{\sin\pi\eta}\cdot
\end{equation}

\section{Cylindrical functions and the exterior solution 
}
\label{asymp}
The asymptotic behavior of 
Bessel functions and their derivatives at infinity 
as $z\rightarrow\infty$ can be found directly in Ref.\ \cite{AS}.
The asymptotic behavior of the Bessel function
$J_\nu(z)$
and their derivatives at the origin as $z\rightarrow 0$ is
determined by relation (9.1.10) of Ref.\ \cite{AS}.
The asymptotic behavior of the Hankel function at the origin 
for $|\nu|<1$ can be calculated from the asymptotic
behavior of $J_\nu(z)$ by using relation
(9.1.3) of Ref.\ \cite{AS},
\begin{eqnarray}
H^{(1)}_\nu(z)&=&\frac{1}{i\sin\nu\pi}\sum_{m=0}^\infty
\frac{(-z^2/4)^m}{m!}\left[\frac{(z/2)^{-\nu}}{\Gamma(1-\nu+m)}
-e^{-\nu\pi i}\,\frac{(z/2)^\nu}{\Gamma(1+\nu+m)}\right],
\\
H^{(1)}_0(z)&=& 1-\frac{2}{\pi i}\left[\ln\left(\frac{z}{2}\right)+
\gamma\right] + {\cal O}(z^2\ln z),
\end{eqnarray}
where $\gamma$ is Euler's constant.
The asymptotic behavior of 
$K_\nu(z)$ for $\nu<1$ when $z\rightarrow 0$
is determined with the help of the last two formulas
and  relations (9.6.4) and (9.6.6) of Ref.\ \cite{AS},
that hold for $-\pi<\arg\, z\leq\pi/2$.
Therefore,
\begin{equation}
K_0(z) \sim -\gamma-\ln (z/2)+{\cal O}(z^2\ln z),
\end{equation}
and
\begin{eqnarray}
K_\nu(z)&\sim&\frac{1}{2\sin\nu\pi}\frac{\pi}{\Gamma(1-\nu)}\left[
\left(\frac{z}{2}\right)^{-\nu}
-\frac{\Gamma(1-\nu)}{\Gamma(1+\nu)}
\left(\frac{z}{2}\right)^\nu\right] +
{\cal O}(z^{2-\nu})\nonumber
\\
&=& \frac{1}{2}\Gamma(\nu)\left[\left(\frac{z}{2}\right)^{-\nu}-
\frac{\Gamma(1-\nu)}{\Gamma(1+\nu)}
\left(\frac{z}{2}\right)^\nu\right]
 +{\cal O}(z^{2-\nu}),
\label{kasymp}
\end{eqnarray}
\begin{eqnarray}
K_\nu'(z)=
-\frac{1}{4}\Gamma(1+\nu)\left[ \left(\frac{z}{2}\right)^{-\nu-1}+
\frac{\Gamma(1-\nu)}{\Gamma(1+\nu)}
\left(\frac{z}{2}\right)^{\nu-1}\right]
 +{\cal O}(z^{1-\nu}),
\end{eqnarray}
otherwise.
Here we have used the identity $\nu\Gamma(\nu)=\Gamma(1+\nu)$ together
with
\begin{equation}
\Gamma(\nu)\Gamma(1-\nu)=\frac{\pi}{\sin\nu\pi}\cdot
\label{gamma1}
\end{equation}

The wave function outside the flux tube is
\begin{equation}
\psi_l(r,\varphi)=K_{|l+\alpha|}(\kappa_l r)e^{il\varphi}.
\end{equation}
For its logarithmic derivative 
 $rK_\nu'(r)/K_\nu(r)$ one has in general 
(see formula 9.6.26 of Ref.\  \cite{AS})
\begin{equation}
z\frac{K_\nu'(z)}{K_\nu(z)}=-\nu -z\frac{K_{\nu-1}(z)}{K_\nu(z)}=
\nu-z\frac{K_{\nu+1}(z)}{K_\nu(z)}\cdot
\end{equation}
Its asymptotic behavior as $z\rightarrow 0$ depends 
whether $\nu$ is less than, equal to, or greater than $1$.
One has
\begin{equation}
z\frac{K_\nu'(z)}{K_\nu(z)}\sim 
-\nu -2\nu\frac{\Gamma(1-\nu)}{\Gamma(1+\nu)}\,
\left(\frac{z}{2}\right)^{2\nu} +{\cal O}(z^{4\nu})
\label{nul1}
\end{equation}
for $0<\nu<1/2$,
\begin{equation}
z\frac{K_\nu'(z)}{K_\nu(z)}\sim 
-\nu -2\nu\frac{\Gamma(1-\nu)}{\Gamma(1+\nu)}\,
\left(\frac{z}{2}\right)^{2\nu} +{\cal O}(z^{2})
\label{nul12}
\end{equation}
for $1/2\leq\nu<1$,
\begin{equation}
z\frac{K_\nu'(z)}{K_\nu(z)}\sim -1 
+ \frac{1}{\Gamma(2)} z^2\ln z +{\cal O}(z^2)
\label{nue1}
\end{equation}
for $\nu=1$, and
\begin{equation}
z\frac{K_\nu'(z)}{K_\nu(z)}\sim -\nu -\frac{1}{2(\nu-1)} z^2 +
{\cal O}(z^{2\nu},\,z^4)
\label{nug1}
\end{equation}
for $\nu>1$.
At infinity as $r\rightarrow \infty$
\begin{equation}
r\frac{K_\nu'(r)}{K_\nu(r)}\sim -r -\frac{1}{2} + {\cal O}(r^{-1}).
\label{inf1}
\end{equation}

\section{Hypergeometric function and the interior solution}
\label{apehyper}
In the case of homogeneous field regularization,
the wave function inside the flux tube of radius $R$
at the energy $E$  is given by (cf. Ref.\ \cite{La}, p. 458)
\begin{eqnarray}
\lefteqn{
\psi({\bf r})=  e^{-\xi/2} \xi^{|l|/2}\hspace*{2cm}}
\nonumber\\
&&
M\left(
\frac{|l|+l}{2}-\frac{\varepsilon(r)}{2}-\frac{1}{2}\,k^2l_B^2,
|l|+1, \xi\right)  e^{il\varphi}.
\label{psiint}
\end{eqnarray}
Here $\xi$ is the flux within the radius $r$ in units
of the flux quantum $\Phi_0$, $\xi=l^2_B r^2/2=\Phi(r)/\Phi_0$,
$l_B=(\hbar c/|e|B)^{1/2}$ is the magnetic length, 
$k^2=2mE/\hbar^2$, and the magnetic moment coupling has been
assumed inside the flux tube. Since we are interested in bound states,
the parameter $x=\kappa R$ will be introduced in analogy with 
Ref.\ \cite{BV} where, as above, $\kappa=\sqrt{2m|E_b|}/\hbar^2$,
$E_b$ being the bound state energy. One can show that
\begin{equation}
 \frac{x^2}{2\alpha} =-\frac{2mE_b}{\hbar^2}l_B^2 .
\end{equation}
The logarithmic derivative of (\ref{psiint}) at the flux tube radius 
$r=R$ is then 
\begin{equation}
\left.r\frac{\psi({\bf r})'}{\psi({\bf r})}
\right|_{r=R} =  -\alpha+|l|+ 2\alpha
\frac{a}{b}\,
\frac{M\left(a+1,b+1,\alpha\right)}
{M\left(a,b,\alpha\right)},
\end{equation}
where $a$ and $b$ are given by Eq. (\ref{abedef}).

The Kummer function  $M(a,b,z)$, frequently denoted by
${}_1F_1(a,b,z)$ \cite{AS}, is defined by
\begin{equation}
M(a,b,z)= 1+ \frac{a}{b}z +\frac{a(a+1)}{b(b+1)}\frac{z^2}{2!} +
\ldots +\frac{a_n}{b_n}\frac{z^n}{n!}+\ldots.
\label{hyperdef}
\end{equation}
where 
\begin{eqnarray}
a_n &=& a (a+1)\ldots (a+n-1),\nonumber\\ 
 b_n &=& b(b+1)\ldots (b+n-1).
\end{eqnarray}
Due to the formula 13.4.8 of Ref.\ \cite{AS}
\begin{equation}
\frac{d}{dz}M(a,b,z)= \frac{a}{b} M(a+1,b+1,z).
\label{hypder}
\end{equation}
Using the simple fact that, provided $b>0$,
\begin{equation}
a\,<\,b \hspace*{0.5cm}\Longrightarrow 
\hspace*{0.5cm}\frac{a}{b}\,<\,\frac{a+1}{b+1},
\end{equation}
one finds for real $z=c> 0$
\begin{equation}
M(a,b,c)<M(a+1,b+1,c).
\label{mrel}
\end{equation}
Note that in the  cases discussed here [see  Eqs. (\ref{abdef}) and 
(\ref{abedef})] one has not only $b>0$ but even $b\geq1$.
Moreover,
provided there is neither magnetic moment coupling nor an attractive
potential inside the flux tube one also has $a>0$ [see Eq. (\ref{abdef})], 
and Eq. (\ref{mmrel}) holds.
However, one cannot maintain
\begin{equation}
M(a,b,c)>0
\end{equation}
when an attractive potential of whatever origin
is  inside the flux tube. For example, for
$g_m-2=2\varepsilon<4$ and 
$x\in\left[0,\sqrt{2\alpha\varepsilon}\,\right)$
 [see Eq. (\ref{abedef})] one has $-1<a<0$. In this case  
$a+1>0$ and $M(a+1,b+1,c)>0$.
From the definition (\ref{hyperdef}) of $M(a,b,c)$ one finds 
easily for such $a$ that
\begin{equation}
M(-|a|,b,c)=2-M(|a|,b,c).
\end{equation}
Therefore, if the relation 
\begin{equation}
M(|a|,b,c) >2
\label{acon}
\end{equation}
is satisfied, one finds  $M(a,b,c)<0$.
Hence, 
\begin{equation}
\frac{M(a+1,b+1,c)}{M(a,b,c)}<0.
\label{ratio1}
\end{equation}
The ratio at $x=0$ is nothing but a parameter $\alpha_1$ of
 Ref. \cite{BV}. It is here where the error \cite{BV1} in 
Ref.\ \cite{BV} is 
made since they claimed $\alpha_1$ to be positive for whatever are
the parameters $a$, $b$, and $c$. For our purposes it is more
 important to discuss the property 
of the ratio given in Eq.\ (\ref{mmrel}) in place of (\ref{ratio1}). 
Note that the ratio stays positive,
\begin{equation}
\frac{2a}{b}\,\frac{M(a+1,b+1,\alpha)}{M(a,b,\alpha)}>0,
\label{ratio2}
\end{equation}
provided Eq.\ (\ref{acon}) holds.
Therefore,  at those values of $x$ [it enters Eq.\ (\ref{acon})
via $a$] it is impossible to satisfy Eq.\ (\ref{matching}) with 
the homogeneous field regularization.

\section{$\zeta$ functions}
\label{ap:zeta}
The Riemann $\zeta$ function is defined by
\begin{equation}
\zeta_R(s) := \sum_{l=1}^\infty l^{-s}.
\end{equation}
It is an analytic function with a simple pole at $s=1$.
The Hurwitz $\zeta_H(s,x)$ function, $x\not\in Z $,
 is a generalization of $\zeta_R(s)$,
\begin{equation}
\zeta_H(s,x) := \sum_{l=0}^\infty (l+x)^{-s}.
\end{equation}       
For $s=-1$
\begin{equation}
\zeta_R(-1)= -\frac{1}{12},
\end{equation}
and 
\begin{equation}
\zeta_H(-1, x) = \frac{1}{2} x(1-x)-\frac{1}{12}=
\frac{1}{2} x(1-x) +\zeta_R(-1).
\end{equation}


\begin{figure}
\centerline{\epsfxsize=7cm \epsfbox{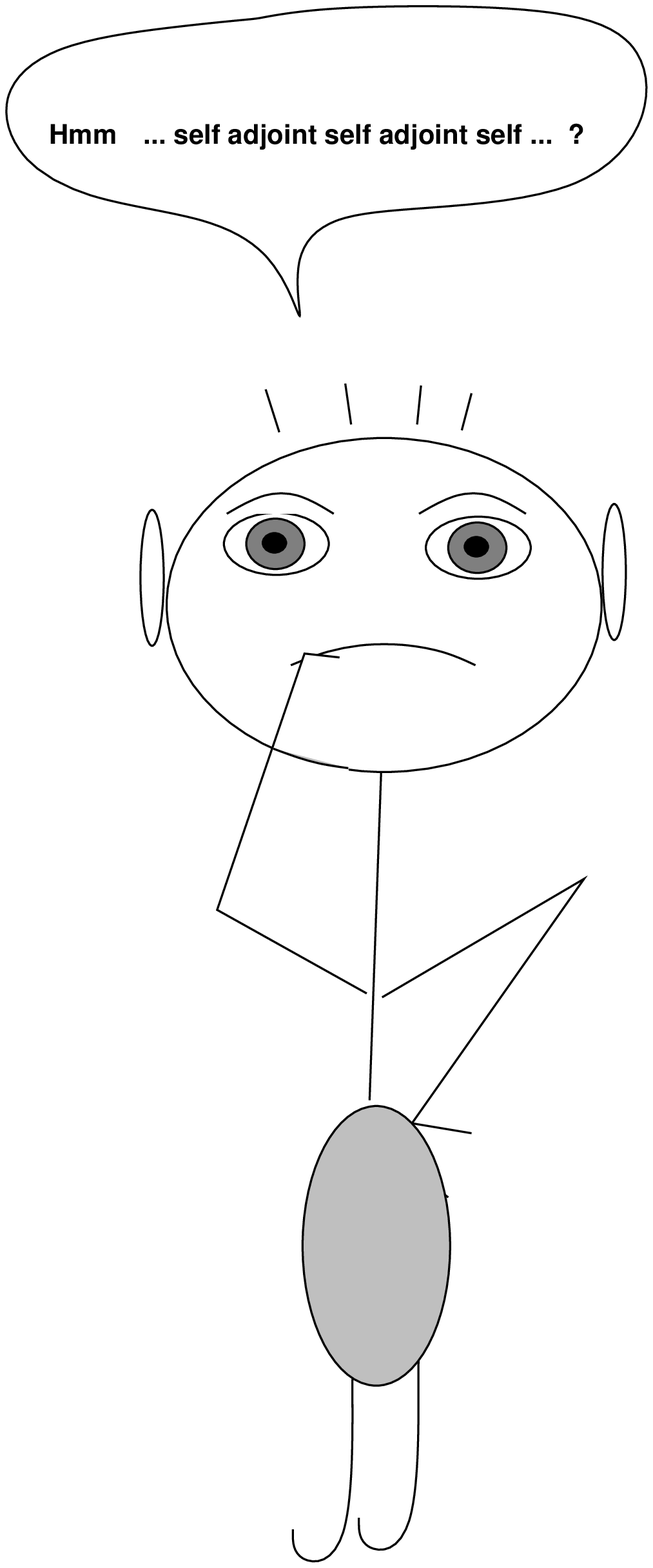}}
\end{figure}

\begin{thebibliography}{99}
\bibitem{AMB}Recently we have shown that the Krein-Friedel formula
can be applied to the Maxwell equations as well. See
A. Moroz, Phys. Rev. B {\bf 51}, 2068 (1995).
\bibitem{Pro}With any self-adjoint operator $H$
one can associate a spectral family $\{{\cal E}(E)\}$. If $f(z)$ is
any complex-valued continuous function, one can define
$f(H):=\int_{-\infty}^\infty f(E)\,d{\cal E}(E)$ (see the book of 
Kato \cite{Ka}, Sec. VI.5.2).
The expansion (\ref{gr:def}) then follows from the fact that
the right derivative $d{\cal E}/dE$ is given by
$$\left.\frac{d{\cal E}(E)}{dE}\right|_{+} 
= -\frac{1}{\pi} \mbox{Im}\,G(E+i\epsilon),$$ and from 
the discussion given below.
\bibitem{Wro}It is worthwhile to mention that if  
$f_1(r)$ and $f_2(r)$ are eigenfunctions of $H_l$ [see Eq.\ (\ref{eigen})]
with respective eigenvalues $k_1^2$ and $k_2^2$ that are different, 
the vanishing
of the Wronskian $W(f_1,f_2)$ at some $r$ does not mean that
 the solutions are linearly dependent.
Indeed, one has $W'(f_1,f_2)=-(1/r)W(f_1,f_2)-(k_1^2-k_2^2)f_1f_2=
-(k_1^2-k_2^2)f_1f_2\neq 0$ at that point and the Wronskian ceases to be zero
at neighboring points.
\bibitem{BV1}We have found that some conclusions of Ref.\ \cite{BV} 
are not valid: their coefficients $\alpha_i$ might be 
negative when they claim them to be
positive; there is an error in calculating the 
number of bound states.
\bibitem{SI}Note that in the absence of bound states
the system enjoys scale invariance.
\end{thebibliography}
\end{document}